\documentclass[hyper]{JHEP3}


\skip\footins = 1\bigskipamount plus 2pt minus 4pt

\usepackage{epsfig}
\newcommand{\be}{\begin{equation}}
\newcommand{\ee}{\end{equation}}
\newcommand{\bea}{\begin{eqnarray}}
\newcommand{\eea}{\end{eqnarray}}

\newcommand{\Slash}[1]{\ooalign{\hfil/\hfil\crcr$#1$}}

\title{Lepton Flavour Violating 
Leptonic/Semileptonic Decays of Charged Leptons in 
the Minimal Supersymmetric Standard Model}

\author{Takeshi Fukuyama$^{\dagger}$, Amon Ilakovac$^{\ddagger}$ 
and Tatsuru Kikuchi$^{\dagger}$\\

$^{\dagger}$ 
Department of Physics, Ritsumeikan University\\ 
Kusatsu, Shiga, 525-8577 Japan\\
E-mail: \email{fukuyama@se.ritsumei.ac.jp},
\email{rp009979@se.ritsumei.ac.jp}\\

$^{\ddagger}$ 
University of Zagreb, Department of Physics\\
P.O. Box 331, Bijeni\v cka cesta 32, HR-10002 Zagreb, Croatia\\
E-mail: \email{ailakov@rosalind.phy.hr}}

\abstract{
We consider the leptonic and semileptonic (SL) lepton flavour violating 
(LFV) decays of the charged leptons in the minimal supersymmetric 
standard model (MSSM). The formalism for evaluation of branching fractions 
for the SL LFV charged-lepton decays with one or two pseudoscalar mesons, 
or one vector meson in the final state, is given. Previous amplitudes 
for the SL LFV charged-lepton decays in MSSM are improved, for instance 
the $\gamma$-penguin amplitude is corrected to assure the gauge invariance. 
The decays are studied not only in the model-independent formulation of 
the theory in the frame of MSSM, but also within the frame of 
the minimal supersymmetric $SO(10)$ model within which the parameters of 
the MSSM are determined. The latter model gives predictions for 
the neutrino-Dirac Yukawa coupling matrix, once free parameters in 
the model are appropriately fixed to accommodate the recent neutrino 
oscillation data. Using this unambiguous neutrino-Dirac Yukawa couplings, 
we calculate the LFV leptonic and SL decay processes 
assuming the minimal supergravity scenario. A very detailed numerical 
analysis is done to constrain the MSSM parameters. 
Numerical results for SL LFV processes are given, for instance for 
$\tau^- \to e^- (\mu^-)\, \pi^0$,
$\tau^- \to e^- (\mu^-)\, \eta$,
$\tau^- \to e^- (\mu^-)\, \eta^\prime$,
$\tau^- \to e^- (\mu^-)\, \rho^0$,
$\tau^- \to e^- (\mu^-)\, \phi$,
$\tau^- \to e^- (\mu^-)\, \omega$, {\it etc.}}

\begin{document}
\section{Introduction}
The neutrino oscillation experiments gave a first experimental evidence 
beyond the standard model (SM) of the electroweak interactions.  
In the SM, neutrinos are massless, purely left-handed  particles, 
so there is no leptonic analogy of the Cabibbo-Kobayashi-Maskawa 
(CKM) matrix in the SM. The neutrino oscillation experiments proved that 
the neutrinos do mix and that they do have mass. The mixing matrix in 
the lepton sector, the Maki-Nakagawa-Sakata (MNS) matrix has bi-large mixing 
structure, indicating that the source of the lepton-flavour mixing is 
different from the corresponding mixing in the quark sector. 
The lepton-flavour mixing observed in neutrino oscillation is the first 
confirmation that the lepton flavour is not conserved quantity. 
Therefore, experimental observation of the other lepton-flavour violating 
(LFV) processes is naturally expected. Theoretical study of such processes 
has a long history before the observation of neutrino oscillations.  
The model independent study of the operators using the SM fields 
\cite{Wilczek:1979hc, Weinberg:1979sa, Weldon:1980gi} shows that 
there are no LFV operators of dimension less or equal to four. 
There is one dimension five LFV operator that induces neutrino oscillations. 
The LFV decays can be induced only with the operators of 
dimension six or more. As the new physics is expected 
to appear at the scale much larger than the electroweak scale 
$\sim 246$ GeV, the LFV decay effects are expected to be much more 
suppressed than the neutrino oscillation effects. Model independent study 
of the LFV processes gives the limits on LFV which every model has to 
satisfy. A model dependent analysis depends on the structure of the model, 
but are much more predictive than the corresponding model independent 
analysis. Therefore, the both approaches are indispensable for a theoretical 
study of LFV. Although the leptonic LFV processes have been studied 
extensively both in model independent way and using various models 
\cite{Hisano:2002zu, Ilakovac:1994kj, Fukuyama:2003hn}, 
the semileptinic (SL) LFV processes have been studied only in few models 
\cite{Ilakovac:1995km, Ilakovac:1995wc, Fajfer:1998px, Ilakovac:1999md, 
Black:2002wh, Sher:2002ew, Brignole:2004ah}. 

The supersymmetric (SUSY) models have much nicer theoretical properties 
than their non-SUSY counterparts. For example, quadratically divergent 
contributions to the Higgs boson mass from heavy (e.g. GUT scale) particles 
cancel with its SUSY partners and as a result the gauge-hierarchy problem 
is much better resolved. The supersymmetrization of the SM cannot be done 
without additional assumptions. For instance, in the supersymmetric version 
of the SM there are dimension four operators violating both the lepton number 
($\Slash{L}$) and the baryon number ($\Slash{B}$), 
leading to very fast proton decay. That led us to the introduction 
of a discrete ${\mathbb{Z}}_2$ symmetry, so called, $R$-parity to forbid such 
undesirable terms. The SUSY breaking also has to be done in such a way 
not to induce too large flavour violation effects. 
There are few successful SUSY breaking mediation mechanisms, such as
gravity mediation \cite{Barbieri:1982eh}, gauge mediation \cite{Dine:1993yw}, 
anomaly mediation \cite{AMSB}, gaugino mediation \cite{gaugino}, 
radion mediation \cite{RMSB}, {\it etc}. 
The best established among them is 
the minimal supergravity model (mSUGRA) \cite{Barbieri:1982eh} which 
assumes that the SUSY breaking occurs in the hidden sector 
at very high scale, 
which communicates with the visible sector (containing SM) with 
flavour-blind gravitational interactions. 
The induced soft SUSY breaking mass terms are requested to be universal 
at the SUSY breaking mediation scale (say, the (reduced) Planck scale), 
and are therefore flavour-diagonal. 
The magnitude of the soft SUSY breaking mass terms obtained is 
in the range to induce potentially observable consequences in the visible sector. 
The renormalization group (RG) flow from the (reduced) Planck scale to 
the mass scale of the right-handed neutrinos, 
induces the flavour non-diagonal 
terms in the SUSY soft breaking terms for the sleptons, through the 
flavor-non diagonal Dirac-neutrino Yukawa matrices they contain \cite{Masiero}.
They can lead to considerable LFV effects, which, depending on the model 
parameters can be in the range of the forthcoming LFV experiments
\cite{Hisano}. 

In this paper, we assume the MSSM with three right-handed 
neutrinos as the low-energy effective theory below the GUT scale. 
In such a framework, neutrino oscillation data suggest the existence of very 
massive right-handed neutrinos which give rise to the small left-handed neutrino 
masses through the see-saw mechanism \cite{see-saw}. In $SO(10)$ models, 
the required right-handed neutrinos can be naturally embedded into 
the common multiplet together with the SM particles for each generation. 
In this paper, the minimal renormazible SUSY $SO(10)$ model 
\cite{Babu, Fukuyama:2002ch, Goh:2003sy, Bajc:2002iw, Dutta:2004wv, 
Matsuda:2004bq} will be taken as a starting theoretical frame. 
One of the advantageous points of this model 
is the automatically conserved $R$-parity 
defined as $R = (-1)^{3 (B-L)+2S}$, where $S$ represents the spin of a field. 
Namely, the $SO(10)$ model discussed here spontaneously breaks the gauged $B-L$ symmetry 
by two units, leading to an automatic $R$-parity conservation. 
The breaking of $SO(10)$ group to the SM gauge group, 
$SU(3)_C \times SU(2)_L \times U(1)_Y$ 
\cite{Fukuyama:2004ps, Fukuyama:2004xs} and its phenomenological consequences 
\cite{Fukuyama:2003hn, Fukuyama:2003uz, Fukuyama:2004pb} 
has already been discussed in our previous publications. 

The main goal of this paper is an analysis of neutrinoless   
SL LFV decays of charged leptons within the MSSM model where parameters are obtained from 
the underlying $SO(10)$ model. At the same time we intend to see how the previous phenomenological 
analyses constrain the LFV parameters. The paper consists basically of 
three parts which are given in three sections. In section two, 
we give the MSSM form factors comprised in the LFV amplitudes 
at the quark-lepton level. We rederive these form factors, 
because some of them were not derived completely in the previous 
literature. In section three, the charged-lepton SL LFV amplitudes at 
lepton-meson level are derived using a simple hadronization procedure 
for the quark currents. The branching ratios corresponding to these 
amplitudes are also given. In section four, the minimal renormalizable 
SUSY $SO(10)$ model is described, the parameters of the MSSM model 
are derived from the $SO(10)$ model, too. Using the predicted Yukawa 
couplings in the minimal $SO(10)$ model, we perform a numerical 
estimation for the SL LFV processes. The last section is devoted 
to a summary. In Appendix A, we give our notation for the mass matrices 
of the neutralino, chargino and sfermions. The MSSM Lagrangian 
for fermion-sfermion-(gaugino, Higgsino) interaction and the trilinear 
interactions with $Z$ boson are given in Appendix B and C, respectively. 
In appendix D we present the loop functions needed to 
evaluate the SL LFV processes. The quark content of meson states, 
essential for the hadronization of quark currents, is listed in 
Appendix E, together with constants that define the hadronized quark 
current in $\gamma$-penguin and $Z$-boson-penguin amplitude. 

\section{Effective Lagrangian for LFV Interactions}
\subsection{Sources for LFV Interactions}
Even though the soft SUSY breaking parameters are flavour blind at 
the scale of the SUSY breaking mediation, the LFV interactions in 
the model can induce the LFV sources at low-energy through 
the renormalization effects. In the following analyses  we assume 
the mSUGRA scenario \cite{Barbieri:1982eh} as the SUSY breaking mediation 
mechanism. At the original scale of the SUSY breaking mediation  
we impose the boundary conditions on the soft SUSY breaking parameters, 
which  are characterized by only five parameters: $m_0$, $M_{1/2}$, 
$A_0$, $B$ and $\mu$. Here, $m_0$ is the universal scalar mass, 
$M_{1/2}$ is the universal gaugino mass, and $A_0$ is the universal 
coefficient of the trilinear couplings.  The parameters in the Higgs 
potential, $B$ and $\mu$, are determined at the electroweak scale 
so that the Higgs doublets obtain the correct electroweak symmetry 
breaking VEV's through the radiative breaking scenario \cite{RBS}. 
The soft SUSY breaking parameters at low energies are obtained through 
the RGE evolutions with the boundary conditions at the GUT scale. 

Although the SUSY breaking mediation scale is normally taken to be 
the (reduced) Planck scale or the string scale ($\sim 10^{18}$ GeV), 
in the following calculations we impose the boundary conditions 
at the GUT scale ($\sim 10^{16}$ GeV), and analyze the RGE evolutions 
from the GUT scale to the electroweak scale. This ansatz is the same 
as the one in the so-called constrained MSSM (CMSSM). 

The effective theory which we analyze below the GUT scale is the MSSM 
with the right-handed neutrinos. The superpotential in the leptonic 
sector is given by 
\begin{eqnarray}
W_Y &=&  Y_{u}^{ij} (u_R^c)_i q_j H_u 
+ Y_d^{ij} (d_R^c)_i q_j H_d 
\nonumber\\
&+& Y_{\nu}^{ij} (\nu_R^c)_i \ell_j H_u 
+ Y_e^{ij} (e_R^c)_i \ell_j H_d 
+ \frac{1}{2} M_{R_{ij}} (\nu_R^c)_i  (\nu_R^c)_j 
+ \mu H_d  H_u  \;  , 
\label{Yukawa4}
\end{eqnarray} 
where the indices $i$, $j$ run over three generations, $H_u$ and $H_d$ 
denote the up-type and down-type MSSM Higgs doublets, respectively, 
and  $M_{R_{ij}}$ is the heavy right-handed Majorana neutrino mass matrix. 
We work in the basis where the charged-lepton Yukawa matrix $Y_e$ and 
the mass matrix $M_{R_{ij}}$ are real-positive and diagonal matrices: 
$Y_e^{ij}=Y_{e_i} \delta_{ij}$ and 
$M_{R_{ij}}=\mbox{diag} ( M_{R_1},  M_{R_2},  M_{R_3}) $. 
Thus, the LFV is originated from the off-diagonal components of 
the neutrino-Dirac-Yukawa coupling matrix $ Y_{\nu}$. 
The soft SUSY breaking terms in the leptonic sector is described as 
\begin{eqnarray}
 -{\cal L}_{\mbox{soft}} 
&=& 
   \tilde{q}^{\dagger}_i 
   \left( m^2_{\tilde{q}} \right)_{ij}
   \tilde{q}_j 
 + \tilde{u}_{R i}^{\dagger} 
   \left( m^2_{ \tilde{u}} \right)_{ij}
    \tilde{u}_{R j} 
 + \tilde{d}_{R i}^{\dagger} 
   \left( m^2_{ \tilde{d}} \right)_{ij} 
\tilde{d}_{R j}    \nonumber  \\ 
\nonumber\\
&+& 
   \tilde{\ell}^{\dagger}_i 
   \left( m^2_{\tilde{\ell}} \right)_{ij}
   \tilde{\ell}_j 
 + \tilde{\nu}_{R i}^{\dagger} 
   \left( m^2_{ \tilde{\nu}} \right)_{ij}
    \tilde{\nu}_{R j} 
 + \tilde{e}_{R i}^{\dagger} 
   \left( m^2_{ \tilde{e}} \right)_{ij} 
\tilde{e}_{R j}    \nonumber  \\ 
&+& m_{H_u}^2 H_u^{\dagger} H_u + m_{H_d}^2 H_d^{\dagger} H_d  
+ \left(  B \mu H_d H_u 
 + \frac{1}{2} B_{\nu} M_{R_{ij}} 
 \tilde{\nu}_{R i}^{\dagger} \tilde{\nu}_{R j} 
+ h.c. \right)  \nonumber \\ 
&+& \left( 
  A_{u}^{ij} \tilde{u}_{R i}^{\dagger}  \tilde{q}_j H_u 
+ A_d^{ij} \tilde{d}_{R i}^{\dagger} \tilde{q}_j H_d  +h.c.  
 \right)  \nonumber \\ 
&+& \left( 
  A_{\nu}^{ij} \tilde{\nu}_{R i}^{\dagger}  \tilde{\ell}_j H_u 
+ A_e^{ij} \tilde{e}_{R i}^{\dagger} \tilde{\ell}_j H_d  +h.c.  
 \right)  \nonumber \\ 
&+& \left( 
    \frac{1}{2} M_1 \tilde{B}  \tilde{B}  
 +  \frac{1}{2} M_2 \tilde{W}^a  \tilde{W}^a  
  + \frac{1}{2} M_3 \tilde{G}^a  \tilde{G}^a  + h.c. \right)  \; .
 \label{softterms} 
\end{eqnarray}
As discussed above, we impose the universal boundary conditions 
at the GUT scale such that 
\begin{eqnarray}
& & \left( m^2_{\tilde{q}} \right)_{ij} 
= \left( m^2_{ \tilde{u}} \right)_{ij}
=   \left( m^2_{ \tilde{d}} \right)_{ij} = m_0^2 \delta_{ij} \; , 
 \nonumber \\ 
& & \left( m^2_{\tilde{\ell}} \right)_{ij} 
= \left( m^2_{ \tilde{\nu}} \right)_{ij}
=   \left( m^2_{ \tilde{e}} \right)_{ij} = m_0^2 \delta_{ij} \; , 
 \nonumber \\ 
& &m_{H_u}^2=m_{H_d}^2 = m_0^2  \; , 
  \nonumber \\ 
& & A_{u}^{ij} = A_0 Y_{u}^{ij}\; , \; \; 
  A_{d}^{ij} = A_0 Y_{d}^{ij} \; , 
  \nonumber \\ 
& & A_{\nu}^{ij} = A_0 Y_{\nu}^{ij}\; , \; \; 
  A_{e}^{ij} = A_0 Y_{e}^{ij} \; , 
  \nonumber \\ 
& & M_1=M_2=M_3= M_{1/2} \; ,  
\end{eqnarray}
and evolve the soft SUSY breaking parameters to the electroweak scale 
according to their RGE's \cite{Martin:1993zk}. The $\mu$ parameter and 
the $B$ parameter are determined at the electroweak scale so as to 
minimize the Higgs potential,
\bea
|\mu|^2 &=& \frac{m_{H_d}^2 - m_{H_u}^2 \tan^2 \beta}{\tan^2 \beta -1} 
- \frac{1}{2}M_Z^2, 
\nonumber\\
B \mu &=& - \frac{1}{2} \left(m_{H_d}^2 + m_{H_u}^2 + 2 |\mu|^2 \right) 
\sin 2 \beta
\left(\ =\ \frac{1}{2} m_A^2 \sin 2\beta \right). 
\eea
The LFV sources in the soft SUSY breaking parameters such as 
the off-diagonal components of $\left( m^2_{\tilde{\ell}} \right)_{ij}$ 
and $A_{e}^{ij}$ are induced by the LFV interactions through the neutrino 
Dirac Yukawa couplings. For example, the LFV effect most directly emerges 
in the left-handed slepton mass matrix through the RGE's such as 
\begin{eqnarray}
\mu \frac{d}{d \mu} 
  \left( m^2_{\tilde{\ell}} \right)_{ij}
&=&  \mu \frac{d}{d \mu} 
  \left( m^2_{\tilde{\ell}} \right)_{ij} \Big|_{\mbox{MSSM}} 
 \nonumber \\
&+& \frac{1}{16 \pi^2} 
\left( m^2_{\tilde{\ell}} Y_{\nu}^{\dagger} Y_{\nu}
 + Y_{\nu}^{\dagger} Y_{\nu} m^2_{\tilde{\ell}} 
 + 2  Y_{\nu}^{\dagger} m^2_{\tilde{\nu}} Y_{\nu}
 + 2 m_{H_u}^2 Y_{\nu}^{\dagger} Y_{\nu} 
 + 2  A_{\nu}^{\dagger} A_{\nu} \right)_{ij}
\label{RGE} 
\end{eqnarray}
where the first term on the right-hand side denotes the MSSM 
term with no LFV. In the leading-logarithmic approximation, 
the off-diagonal components ($i \neq j$) of the left-handed 
slepton mass matrix are estimated as 
\begin{eqnarray}
 \left(\Delta  m^2_{\tilde{\ell}} \right)_{ij}
 \sim - \frac{3 m_0^2 + A_0^2}{8 \pi^2} 
 \left( Y_{\nu}^{\dagger} L Y_{\nu} \right)_{ij} \; ,  
 \label{leading}
\end{eqnarray}
where the distinct thresholds for the right-handed Majorana neutrinos 
are taken into account by the matrix 
$ L_{ij} = \log \left(\frac{M_G}{M_{R_i}} \right) \delta_{ij}$. 
We can see that the neutrino-Dirac-Yukawa coupling matrix plays 
the crucial role in calculations of the LFV processes. 

\subsection{Effective Lagrangian in terms of quark fields and LFV form factors}
In any model containing standard model as the low-energy effective theory, 
an effective Lagrangian for the SL LFV decays of a lepton contains 
only three terms: the photon-penguin, the Z-boson-penguin and the box term, 
\be
i{\cal L}_{\mathrm{eff}}
\left(\ell_i \to \ell_j + \bar{q} + q' \right)
= 
i{\cal L}^{\gamma}_{\mathrm{eff}} 
+ i{\cal L}^{Z}_{\mathrm{eff}} 
+ i{\cal L}^{\mathrm{box}}_{\mathrm{eff}}. 
\ee
These terms have the following generic structure,
\bea
i{\cal L}^{\gamma}_{\mathrm{eff}}(x)
&=&
-i e^2\:  \int\: d^4 y\: \bar{\ell}_j(x)
\left[(- \partial_x^2 \gamma_\mu 
+ {\Slash{\partial}}_x \partial_{x\mu} ) D(x-y)
({\cal P}_{1\gamma}^{L} P_L+ {\cal P}_{1\gamma}^{ R} P_R)
\right.
\nonumber\\
&+&  
\left.
\sigma_{\mu\nu} \partial_x^\nu D(x-y)
({\cal P}_{2\gamma}^{L} P_L + {\cal P}_{2\gamma}^{R} P_R) \right] \ell_i (x)
\nonumber\\ 
&\times& \sum_{q=u,d,s} Q_q \bar{q}(y) \gamma^\mu q(y),
\\
i{\cal L}^{Z}_{\mathrm{eff}} (x) &=&
 i \frac{g^2}{m_Z^2c_W^2}\  \bar{\ell}_j (x) \gamma_\mu 
 ({\cal P}_{Z}^L P_L+{\cal P}_{Z}^R P_R) \ell_i (x)
\nonumber\\ 
&\times&
\sum_{q=u,d,s} \bar{q} (I_{3q}-2Q_q s_W^2)\gamma^\mu 
- I_{3q} \gamma^\mu\gamma_5) q(x),
\\
i{\cal L}^{\mathrm{box}}_{\mathrm{eff}} &=& i \,
\sum_{\bar{q}_aq_b=\bar{u}u,\bar{d}d,\bar{s}s,\bar{d}s,\bar{s}d}
\bigg[{\cal{B}}_{1 \bar{q}_a q_b}^L 
(\bar{\ell}_j \gamma^\mu P_L \ell_i )
(\bar{q}_a \gamma_\mu P_L q_b )
+{\cal{B}}_{1 \bar{q}_a q_b}^R 
(\bar{\ell}_j \gamma^\mu P_R \ell_i ) 
(\bar{q}_a\gamma_\mu P_R q_b )
\nonumber\\
&+& {\cal{B}}_{2 \bar{q}_a q_b}^L 
(\bar{\ell}_j \gamma^\mu P_L \ell_i ) 
(\bar{q}_a \gamma_\mu P_R q_b ) 
+{\cal{B}}_{2 \bar{q}_a q_b}^R 
(\bar{\ell}_j \gamma^\mu P_R \ell_i ) 
(\bar{q}_a \gamma_\mu P_L q_b )
\nonumber\\
&+& {\cal{B}}_{3 \bar{q}_a q_b}^L 
(\bar{\ell}_j P_L \ell_i )
(\bar{q}_a P_L q_b )
+{\cal{B}}_{3 \bar{q}_a q_b}^R 
(\bar{\ell}_j P_R \ell_i )
(\bar{q}_a P_R q_b )
\nonumber\\
&+& \bar{{\cal{B}}}_{3 \bar{q}_a q_b}^L 
(\bar{\ell}_j P_L \ell_i )
(\bar{q}_a P_R q_b )
+\bar{{\cal{B}}}_{3 \bar{q}_a q_b}^R 
(\bar{\ell}_j P_R \ell_i )
(\bar{q}_a P_L q_b )
\nonumber\\
&+& {\cal{B}}_{4 \bar{q}_a q_b}^L 
(\bar{\ell}_j \sigma_{\mu\nu} P_L \ell_i) 
(\bar{q}_a \sigma^{\mu\nu} P_R q_b )
+ {\cal{B}}_{4 \bar{q}_a q_b}^R 
(\bar{\ell}_j \sigma_{\mu\nu} P_R \ell_i ) 
(\bar{q}_a \sigma^{\mu\nu} P_L q_b )
\bigg], 
\eea
where $s_W=\sin\theta_W$ and $c_W=\cos\theta_W$, $Q_q$ is the quark charge 
in units of $e$, and $I_{3q}$ is weak quark isospin. $D(x-y)$ is the Green 
function for the massless scalar particle, contained in the photon propagator. 
The structure of the photon-penguin term in the effective Lagrangian is 
a consequence of the gauge invariance. Especially, the first term must 
contain ${\Slash{\partial}}_x \partial_{x\mu}$, which was neglected in 
reference \cite{Hisano}. The information on the model under consideration 
is contained in the form factors ${\cal P}^{L,R}_{a\gamma}$, $a=1,2$, 
${\cal P}^{L,R}_Z$, $\bar{{\cal{B}}}_{3 \bar{q}_a q_b}^L$, and 
${\cal B}_{a \bar{q}_a q_b}^{L,R}$, $a=1,2,3,4$. In the following 
three subsections these form factors are given for the MSSM. 

\subsubsection{The photon-penguin form factors}
The amplitude for $\ell_i\to \ell_j \gamma^*$ for an off-mass-shell 
photon process is obtained the from corresponding part of the effective 
Lagrangian neglecting the quark current and the photon propagator,
\be
{\cal M_\mu}^\gamma = i\: T_\mu^\gamma \ =\ 
-e \overline{u}_{\ell_j}
\bigg[
(q^2\gamma_\mu - q_\mu \Slash{q}) 
({\cal P}_{1\gamma}^{L} P_L + {\cal P}_{1\gamma}^{R} P_R)
\ +\
i\sigma_{\mu\nu} q^\nu 
({\cal P}_{2\gamma}^{L} P_L + {\cal P}_{2\gamma}^{R} P_R)
\bigg] u_{\ell_i}.
\ee
The amplitude is written without photon polarization vector. 

In the MSSM the photon-penguin amplitude has two contributions, 
a chargino and a neutralino contribution. That reflects 
in the structure of the form factors, 
\begin{eqnarray}
{\cal P}_{a\gamma}^{L,R}
&=&{\cal P}_{a\gamma}^{(C)L,R}
+{\cal P}_{a\gamma}^{(N)L,R},\qquad\ a=1,2 
\end{eqnarray}
with $C$ and $N$ subscript denoting the chargino and neutralino 
part of a form factor. 

Because of the gauge invariance, the zeroth-order and the first-order term 
in Taylor expansion in momenta and masses of incoming and outgoing 
particles are equal zero. Here, the second order term in Taylor 
expansion is presented, and higher order terms are neglected. 

The neutralino contributions are 
\bea
{\cal P}_{1\gamma}^{(N)L} &=& 
\frac{i}{576\pi^2} N^{R(e)}_{jAX}  N^{R(e)*}_{iAX} 
\frac{1}{m^2_{\tilde{e}_X}}
\nonumber\\
&\times&
\frac{11(x_{AX}^0)^3 - 18(x_{AX}^0)^2 + 9(x_{AX}^0) 
- 2 - 6(x_{AX}^0)^3\ln(x_{AX}^0)}{(1-x_{AX}^0)^4},
\\
{\cal P}_{1\gamma}^{(N)R} 
&=&{{\cal P}_{1\gamma}^{(N)L}}|_{L \leftrightarrow R}, 
\\
{\cal P}_{2\gamma}^{(N)L} &=&
\frac{i}{32\pi^2} 
\Bigg[
N^{R(e)}_{jAX}  N^{R(e)*}_{iAX}
(-1) m_j \frac{1}{m^2_{\tilde{e}_X}}
\nonumber\\
&\times&
\frac{2(x_{AX}^0)^3 + 3(x_{AX}^0)^2 - 6(x_{AX}^0) 
+ 1 - 6(x_{AX}^0)^2\ln(x_{AX}^0)}{6(1-x_{AX}^0)^4}
\nonumber\\
&+&
N^{L(e)}_{jAX}  N^{L(e)*}_{iAX}
(-1) m_i \frac{1}{m^2_{\tilde{e}_X}}
\nonumber\\
&\times&
\frac{2(x_{AX}^0)^3 + 3(x_{AX}^0)^2 - 6(x_{AX}^0) 
+ 1 - 6(x_{AX}^0)^2\ln(x_{AX}^0)}{6(1-x_{AX}^0)^4}
\nonumber\\
&+&
N^{L(e)}_{jAX}  N^{R(e)*}_{iAX}
(-1) m_{\tilde{\chi}^0_A} \frac{1}{m^2_{\tilde{e}_X}}
\frac{-(x_{AX}^0)^2 + 1 + 2(x_{AX}^0)\ln(x_{AX}^0)}{(1-x_{AX}^0)^3}
\Bigg], 
\\
{\cal P}_{2\gamma}^{(N)R} 
&=&{{\cal P}_{2\gamma}^{(N)L}}|_{L \leftrightarrow R}, 
\eea
where $x_{AX}^{0}={M_{\tilde{\chi}_A^{0}}^2}/{m_{\tilde{e}_X}^2}$. 
The chargino contributions are given by 
\bea
{\cal P}_{1\gamma}^{(C)L} &=&
\frac{i}{576\pi^2} C^{R(e)}_{jAX}  C^{R(e)*}_{iAX}
\frac{1}{m^2_{\tilde{\nu}_X}}
\nonumber\\
&\times&
\frac{16 - 45(x_{AX}^-) + 36(x_{AX}^-)^2 - 
7(x_{AX}^-)^3 + 6(2-3(x_{AX}^-))\ln(x_{AX}^-)}{(1-x_{AX}^-)^4},
\\
{\cal P}_{1\gamma}^{(C)R} &=&
{{\cal P}_{1\gamma}^{(C)L}}|_{L \leftrightarrow R},
\\
{\cal P}_{2\gamma}^{(C)L} &=&
\frac{i}{32\pi^2}
\Bigg[
C^{R(e)}_{jAX}  C^{R(e)*}_{iAX}m_j \frac{1}{m^2_{\tilde{\nu}_X}}
\nonumber\\
&\times&
\frac{2 + 3(x_{AX}^-) - 6(x_{AX}^-)^2 + (x_{AX}^-)^3 
+ 6(x_{AX}^-)\ln(x_{AX}^-)}{6(1-x_{AX}^-)^4}
\nonumber\\
&+&
C^{L(e)}_{jAX}  C^{L(e)*}_{iAX}
 m_i \frac{1}{m^2_{\tilde{\nu}_X}}
\nonumber\\
&\times&
\frac{2 + 3(x_{AX}^-) - 6(x_{AX}^-)^2 
+ (x_{AX}^-)^3 + 6(x_{AX}^-)\ln(x_{AX}^-)}{6(1-x_{AX}^-)^4}
\nonumber\\
&+&
C^{L(e)}_{jAX}  C^{R(e)*}_{iAX}
 m_{\tilde{\chi}^-_A} \frac{1}{m^2_{\tilde{\nu}_X}}
\frac{-3 + 4(x_{AX}^-) - (x_{AX}^-)^2 - 2\ln(x_{AX}^-)}
 {(1-x_{AX}^-)^3}
\Bigg], 
\\
{\cal P}_{2\gamma}^{(C)R} &=&
{{\cal P}_{2\gamma}^{(C)L}}|_{L \leftrightarrow R}, 
\eea
where $x_{AX}^{-}={M_{\tilde{\chi}_A^{-}}^2}/{m_{\tilde{\nu}_X}^2}$. 
The form factor contributions are written in the same way as in 
Ref. \cite{Hisano}, including the loop functions into the expressions, 
to make the comparison with the results of Ref. \cite{Hisano} easy. 
Both chargino and neutralino parts of the form factors agree with 
the corresponding form factors in the reference \cite{Hisano} if 
the terms proportional to the mass of the lighter mesons $m_j$ are 
neglected. Nevertheless, these terms cannot be neglected, because 
the constants $N^{L,R(e)}_{jAX}$ and $C^{L,R(e)}_{jAX}$ (see Appendix B) 
also depend on the lepton masses is such a way that in some cases 
the term proportional to $m_j$ is larger than the term proportional 
to the mass of decaying lepton $m_i$. 

\subsubsection{The Z-penguin form factors}
The amplitude for $\ell_i\to \ell_j Z^*$, for the off-mass-shell 
Z-boson process, 
(obtained analogously as the photon-penguin amplitude) is 
\bea
{\cal M}^Z_\mu 
&=& 
\frac{i}{(4\pi)^2} g \overline{u}_{l_j}
\bigg[
\gamma_\mu P_L ({\cal P}_{Z}^{(C)L}
+{\cal P}^{(N)L}_{Z})
+
\gamma_\mu P_R ({\cal P}^{(C)R}_{Z}
+{\cal P}^{(N)R}_{Z})
\bigg]
u_{l_i},
\eea
where ${\cal P}^{(C){L,R}}_{Z}$ and ${\cal P}^{(N){L,R}}_{Z}$ are 
the chargino and neutralino part of the total form factors, 
${\cal P}^{L,R}_{Z}$. Here are the expressions for these form factors: 
\bea
{\cal P}^{(C)L}_{Z} &=&
\label{FZLC}
C^{R(e)}_{jBX}  C^{R(e)*}_{iAX} 
\Big[
E^{L(\tilde{\chi}^-)}_{BA} m_{\tilde{\chi}^-_B} m_{\tilde{\chi}^-_A} 
 F_1(m^2_{\tilde{\nu}_X}, m^2_{\tilde{\chi}_A^-}, m^2_{\tilde{\chi}_B^-})
\nonumber\\
&-& 2 E^{R(\tilde{\chi}^-)}_{BA} 
F_2(m^2_{\tilde{\nu}_X}, m^2_{\tilde{\chi}_A^-}, m^2_{\tilde{\chi}_B^-})
+
\delta_{AB} G_{Ze}^L f_1 (m^2_{\tilde{\nu}_X}, m^2_{\tilde{\chi}_A^-})
\Big]
\nonumber\\
&+&\Big\{
C^{R(e)}_{jBX} E_{BA}^{L(\tilde{\chi}^-)}  C^{L(e)*}_{iAX}
[-2 F_2(m^2_{\tilde{\nu}_X}, m^2_{\tilde{\chi}_A^-}, m^2_{\tilde{\chi}_B^-})
]
\Big\},
\\
{\cal P}^{(N)L}_{Z} &=&
\label{FZLN}
N^{R(e)}_{jBY}  N^{R(e)*}_{iAX}
\Big[
- 2 D_{YX}^{(\tilde{e})} 
 F_2(m^2_{\tilde{\chi}_A^0}, m^2_{\tilde{e}_X}, m^2_{\tilde{e}_Y})
+
\delta_{YX} G_{Ze}^L f_2 (m^2_{\tilde{\chi}_A^0}, m^2_{\tilde{e}_X})
\Big]
\nonumber\\
&+&
\Big\{
N^{L(e)}_{jBX} E_{BA}^{L(\tilde{\chi}^0)}  N^{R(e)*}_{iAX}
[
m_{\tilde{\chi}^0_B} m_{\tilde{\chi}^0_A}
 F_1(m^2_{\tilde{e}_X}, m^2_{\tilde{\chi}_A^0}, m^2_{\tilde{\chi}_B^0})
]
\nonumber\\
&+&
N^{R(e)}_{jBX} E_{BA}^{L(\tilde{\chi}^0)}  N^{L(e)*}_{iAX}
[
-2 F_2(m^2_{\tilde{e}_X}, m^2_{\tilde{\chi}_A^0}, m^2_{\tilde{\chi}_B^0})  
]
\Big\},
\\
{\cal P}^{(C)R}_{Z} &=& {\cal P}^{(C)L}_{Z} (L\leftrightarrow R),
\\
{\cal P}^{(N)R}_{Z} &=& {\cal P}^{(N)L}_{Z} (L\leftrightarrow R). 
\mbox{\hspace{1cm}} 
E_{BA}^{R(\tilde{\chi}^0)}\ =\ -E_{BA}^{L(\tilde{\chi}^0)}. 
\eea
$G_{Ze}^L$ and $G_{Ze}^R$ are constants appearing in 
the SM $Ze_ie_j$ vertices, 
\bea
{\cal L}_{Ze_ie_j}&=&
-g\gamma_\mu\delta_{ij}\{G_{Ze}^L P_L + G_{Ze}^R P_R\}
\nonumber\\
&&-g\gamma_\mu\delta_{ij}\left\{
\left[-\frac{1}{2}\frac{1}{c_W}+\frac{s^2_W}{c_W}\right]P_L
+\left[\frac{s^2_W}{c_W}\right]P_R\right\},
\eea
while $E_{BA}^{R(\tilde{\chi}^-)}$ and 
$E_{BA}^{L,R(\tilde{\chi}^0)}$ are constants 
in the $Z$-boson---chargino and $Z$-boson---neutralino vertices, and 
$D_{YX}^{(\tilde{e})}$ is a constant in the $Z$-boson---selectron vertex. 
These constants are defined in Appendix C. 
$F_1(a,b,c)$ and $F_2(a,b,c)$ are loop functions contained in 
the triangle-diagram part of the amplitude, and $f_1$ and $f_2$ are 
the loop functions coming from the self-energy part of the amplitude. 
They are given in Appendix D.

The terms in Eqs. (\ref{FZLC}) and (\ref{FZLN}) which have a corresponding 
contribution in the photon amplitude (the leading order photon-penguin 
amplitude comes from six Feynman diagrams, while the $Z$-boson-penguin 
amplitude has eight Feynman-diagram contributions) have been compared by 
replacing $Z$-boson vertices with corresponding photon vertices, and 
an agreement was found. The remaining two Feynman diagram contributions 
which are embraced by curly brackets in Eqs. (\ref{FZLC}) and (\ref{FZLN}) 
have been checked carefully. The new terms in ${\cal P}^{(C)L}_{Z}$ 
in comparison with Ref. \cite{Hisano} are third (self-energy--type term) 
and fourth term. Further, neither of our terms in ${\cal P}^{(C)R}_{Z}$ 
does agree with amplitude in Ref. \cite{Hisano}, although the expression 
in the curly brackets is almost equal to it. 

\subsubsection{The box form factors}
The box contribution to the SL LFV $\ell\to \ell_i \bar{q}_a q_b$ 
amplitude comes from two box-diagrams in the leading order of perturbation 
theory. The box-amplitude reads
\bea
\label{Mbox}
{\cal M}_{\mathrm{box}} &=& 
\frac{i}{(4 \pi)^2}
\sum_{\bar{q}_a q_b=\bar{u}u,\bar{d}d,\bar{s}s,\bar{d}s,\bar{s}d} 
\bigg[{\cal{B}}_{1 \bar{q}_a q_b}^L 
(\bar{u}_{\ell_j} \gamma^\mu P_L u_{\ell_i}) 
(\bar{u}_{q_a} \gamma_\mu P_L v_{q_b})
+{\cal{B}}_{1 \bar{q}_a q_b}^R 
(\bar{u}_{\ell_j} \gamma^\mu P_R u_{\ell_i}) 
(\bar{u}_{q_a} \gamma_\mu P_R v_{q_b})
\nonumber\\
&+&
{\cal{B}}_{2 \bar{q}_a q_b}^L 
(\bar{u}_{\ell_j} \gamma^\mu P_L u_{\ell_i}) 
(\bar{u}_{q_a} \gamma_\mu P_R v_{q_b}) 
+{\cal{B}}_{2 \bar{q}_a q_b}^R 
(\bar{u}_{\ell_j} \gamma^\mu P_R u_{\ell_i}) 
(\bar{u}_{q_a} \gamma_\mu P_L v_{q_b})
\nonumber\\
&+&
{\cal{B}}_{3 \bar{q}_a q_b}^L 
(\bar{u}_{\ell_j} P_L u_{\ell_i}) 
(\bar{u}_{q_a} P_L v_{q_b})
+{\cal{B}}_{3 \bar{q}_a q_b}^R 
(\bar{u}_{\ell_j} P_R u_{\ell_i}) 
(\bar{u}_{q_a} P_R v_{q_b})
\nonumber\\
&+&
\bar{{\cal{B}}}_{3 \bar{q}_a q_b}^L
(\bar{u}_{\ell_j} P_L u_{\ell_i}) 
(\bar{u}_{q_a} P_R v_{q_b})
+ \bar{{\cal{B}}}_{3 \bar{q}_a q_b}^R
(\bar{u}_{\ell_j} P_R u_{\ell_i}) 
(\bar{u}_{q_a} P_L v_{q_b})
\nonumber\\
&+&
{\cal{B}}_{4 \bar{q}_a q_b}^L 
(\bar{u}_{\ell_j} \sigma_{\mu\nu} P_L u_{\ell_i}) 
(\bar{u}_{q_a} \sigma^{\mu\nu} P_L v_{q_b})
\nonumber\\
&+&
{\cal{B}}_{4 \bar{q}_a q_b}^R 
(\bar{u}_{\ell_j} \sigma_{\mu\nu} P_R u_{\ell_i}) 
(\bar{u}_{q_a} \sigma^{\mu\nu} P_R v_{q_b})
\bigg].
\eea
The very rich structure of the box-diagram amplitude is a consequence 
of the Fiertz transformation of the terms with product of lepton-quark 
and quark-lepton vector and axial-vector currents. All currents permitted 
by the Dirac algebra do appear. Each box-amplitude form factor has 
a chargino (C) and a neutralino (N) contribution.
\begin{eqnarray}
{\cal{B}}^{L,R}_{i \bar{q}_a q_b}&=&
{\cal{B}}^{(N)L, R}_{i \bar{q}_a q_b} 
+ {\cal{B}}^{(C)L, R}_{i \bar{q}_a q_b},
\\
\bar{{\cal{B}}}^{L, R}_{3 \bar{q}_a q_b}&=&
\bar{{\cal{B}}}^{(N)L, R}_{3 \bar{q}_a q_b} 
+ \bar{{\cal{B}}}^{(C)L, R}_{3 \bar{q}_a q_b}.
\end{eqnarray}
Here and in the following equations $\bar{q}_aq_b$ assume the values 
appearing in the sum in Eq. (\ref{Mbox}). Neutralino contributions read
\begin{eqnarray}
{\cal{B}}_{1\bar{q}_aq_b}^{(N)L} &=& \frac{1}{4} 
d_2 (M_{\tilde{\chi}_A^0}^2,M_{\tilde{\chi}_B^0}^2,
m_{\tilde{e}_X}^2,m_{\tilde{q}_Y}^2 )
N^{R(e)*}_{iAX}N^{R(e)}_{jBX}
N^{R(q)*}_{bBY}N^{R(q)}_{aAY}
\nonumber \\
&+& \frac{1}{2} 
d_0(M_{\tilde{\chi}_A^0}^2,M_{\tilde{\chi}_B^0}^2,
m_{\tilde{e}_X}^2,m_{\tilde{q}_Y}^2) 
M_{\tilde{\chi}_A^0} M_{\tilde{\chi}_B^0}
N^{R(e)*}_{iAX}N^{R(e)*}_{jBX}
N^{R(q)}_{bAY} N^{R(q)}_{aBY},
\\
{\cal{B}}_{2\bar{q}_aq_b}^{(N)L} &=& 
-\frac{1}{4} d_2(M_{\tilde{\chi}_A^0}^2,M_{\tilde{\chi}_B^0}^2,
m_{\tilde{e}_X}^2,m_{\tilde{q}_Y}^2)
N^{R(e)*}_{iAX}N^{R(e)}_{jBX}
N^{L(q)*}_{bAY} N^{L(q)}_{aBY}
\nonumber \\
&-&\frac{1}{2} 
d_0(M_{\tilde{\chi}_A^0}^2,M_{\tilde{\chi}_B^0}^2,
m_{\tilde{e}_X}^2,m_{\tilde{q}_Y}^2) 
M_{\tilde{\chi}_A^0} M_{\tilde{\chi}_B^0}
N^{R(e)*}_{iAX}N^{R(e)}_{jBX}
N^{L(q)*}_{bAY}N^{L(q)}_{aBY},
\\
{\cal{B}}_{3\bar{q}_aq_b}^{(N)L}&=& 
d_0(M_{\tilde{\chi}_A^0}^2,M_{\tilde{\chi}_B^0}^2,
m_{\tilde{e}_X}^2,m_{\tilde{q}_Y}^2) 
M_{\tilde{\chi}_A^0} M_{\tilde{\chi}_B^0}
\Big\{
-\frac{1}{2}
N^{R(e)*}_{iAX} N^{L(e)}_{jBX}
N^{R(q)*}_{bBY} N^{L(q)}_{aAY}
\nonumber \\ 
&-&\frac{1}{2} 
N^{R(e)*}_{iAX} N^{L(e)}_{jBX}
N^{R(q)*}_{bAY} N^{L(q)}_{aBY}
\Big\},
\\
\bar{{\cal{B}}}_{3\bar{q}_aq_b}^{(N)L}&=& 
d_2(M_{\tilde{\chi}_A^0}^2,M_{\tilde{\chi}_B^0}^2,
m_{\tilde{e}_X}^2,m_{\tilde{q}_Y}^2) 
\Big\{
-\frac{1}{2}
N^{R(e)*}_{iAX} N^{L(e)}_{jBX}
N^{L(q)*}_{bBY} N^{R(q)}_{aAY}
\nonumber \\ 
&-&\frac{1}{2} 
N^{R(e)*}_{iAX} N^{L(e)}_{jBX}
N^{L(q)*}_{bAY} N^{R(q)}_{aBY}
\Big\},
\\
{\cal{B}}_{4\bar{q}_a q_b}^{(N)L}&=& 
\frac{1}{8} d_0(M_{\tilde{\chi}_A^0}^2,M_{\tilde{\chi}_B^0}^2,
m_{\tilde{e}_X}^2,m_{\tilde{q}_Y}^2) 
M_{\tilde{\chi}_A^0}M_{\tilde{\chi}_B^0}
\left\{
N^{R(e)*}_{iAX} N^{L(e)}_{jBX} 
N^{R(q)*}_{bBY} N^{L(q)}_{aAY}
\right.
\nonumber\\
&-&
\left.
N^{R(e)*}_{iAX} N^{L(e)}_{jBX}
N^{R(q)*}_{bAY} N^{L(q)}_{aBY}
\right\},
\\
{\cal{B}}_{i\bar{q}_a q_b}^{(N)R}&=&
{\cal{B}}_{i\bar{q}_a q_b}^{(N)L}|_{L \leftrightarrow R}
~~~~~(i=1, \cdots, 4)
\\
\bar{{\cal{B}}}_{3\bar{q}_a q_b}^{(N)R}&=&
\bar{{\cal{B}}}_{3\bar{q}_a q_b}^{(N)L}|_{L \leftrightarrow R}. 
\end{eqnarray}
The chargino contributions are
\begin{eqnarray}
{\cal{B}}_{1\bar{q}_aq_b}^{(C)L} &=& \frac{1}{4}
d_2 (M_{\tilde{\chi}_A^-}^2,M_{\tilde{\chi}_B^-}^2,
m_{\tilde{\nu}_X}^2,m_{\tilde{q'}_Y}^2 )
C^{R(e)*}_{iAX}C^{R(e)}_{jBX}
C^{R(q)*}_{bBY}C^{R(q)}_{aAY}\ \delta_{qd}
\nonumber \\
&+& \frac{1}{2}
d_0(M_{\tilde{\chi}_A^-}^2,M_{\tilde{\chi}_B^-}^2,
m_{\tilde{\nu}_X}^2,m_{\tilde{q'}_Y}^2)
M_{\tilde{\chi}_A^-} M_{\tilde{\chi}_B^-}
C^{R(e)*}_{iAX}C^{R(e)*}_{jBX}
C^{R(q)}_{bAY} C^{R(q)}_{aBY}\ \delta_{qu},
\\
{\cal{B}}_{2\bar{q}_aq_b}^{(C)L} &=&
-\frac{1}{4} d_2(M_{\tilde{\chi}_A^-}^2,M_{\tilde{\chi}_B^-}^2,
m_{\tilde{\nu}_X}^2,m_{\tilde{q'}_Y}^2)
C^{R(e)*}_{iAX}C^{R(e)}_{jBX}
C^{L(q)*}_{bAY} C^{L(q)}_{aBY}\ \delta_{qd}
\nonumber \\
&-&\frac{1}{2}
d_0(M_{\tilde{\chi}_A^-}^2,M_{\tilde{\chi}_B^-}^2,
m_{\tilde{\nu}_X}^2,m_{\tilde{q'}_Y}^2)
M_{\tilde{\chi}_A^-} M_{\tilde{\chi}_B^-}
C^{R(e)*}_{iAX}C^{R(e)}_{jBX}
C^{L(q)*}_{bAY}C^{L(q)}_{aBY}\ \delta_{qu},
\\
{\cal{B}}_{3\bar{q}_aq_b}^{(C)L}&=&
d_0(M_{\tilde{\chi}_A^-}^2,M_{\tilde{\chi}_B^-}^2,
m_{\tilde{\nu}_X}^2,m_{\tilde{q'}_Y}^2)
M_{\tilde{\chi}_A^-} M_{\tilde{\chi}_B^-}
\Big\{
-\frac{1}{2}
C^{R(e)*}_{iAX} C^{L(e)}_{jBX}
C^{R(q)*}_{bBY} C^{L(q)}_{aAY}\ \delta_{qd}
\nonumber \\
&-&\frac{1}{2}
C^{R(e)*}_{iAX} C^{L(e)}_{jBX}
C^{R(q)*}_{bAY} C^{L(q)}_{aBY}\ \delta_{qu}
\Big\},
\\
\bar{{\cal{B}}}_{3\bar{q}_aq_b}^{(C)L}&=&
d_2(M_{\tilde{\chi}_A^-}^2,M_{\tilde{\chi}_B^-}^2,
m_{\tilde{\nu}_X}^2,m_{\tilde{q'}_Y}^2)
\Big\{
-\frac{1}{2}
C^{R(e)*}_{iAX} C^{L(e)}_{jBX}
C^{L(q)*}_{bBY} C^{R(q)}_{aAY}\ \delta_{qd}
\nonumber \\
&-&\frac{1}{2}
C^{R(e)*}_{iAX} C^{L(e)}_{jBX}
C^{L(q)*}_{bAY} C^{R(q)}_{aBY}\ \delta_{qu}
\Big\},
\\
{\cal{B}}_{4\bar{q}_a q_b}^{(C)L}&=&
\frac{1}{8} d_0(M_{\tilde{\chi}_A^-}^2,M_{\tilde{\chi}_B^-}^2,
m_{\tilde{\nu}_X}^2,m_{\tilde{q'}_Y}^2)
M_{\tilde{\chi}_A^-}M_{\tilde{\chi}_B^-}
\left\{
C^{R(e)*}_{iAX} C^{L(e)}_{jBX}
C^{R(q)*}_{bBY} C^{L(q)}_{aAY}\ \delta_{qd}
\right.
\nonumber\\
&-&
\left.
C^{R(e)*}_{iAX} C^{L(e)}_{jBX}
C^{R(q)*}_{bAY} C^{L(q)}_{aBY}\ \delta_{qu}
\right\},
\\
{\cal{B}}_{i\bar{q}_a q_b}^{(C)R}&=&
{\cal{B}}_{i\bar{q}_a q_b}^{(C)L}|_{L \leftrightarrow R}
~~~~~(i=1, \cdots, 4)
\end{eqnarray}
where for $q = u(d)$, $q' = d(u)$. The sum over paired indices is assumed.
The Kronecker function $\delta_{qu}$ [$\delta_{qu}$] denotes that $q$ quark 
is one of up ($u$, $c$ or $t$) quarks [one of the down quarks].
The loop functions $d_0$ and $d_2$ are evaluated by neglecting the momenta 
of incoming and outgoing particles. They are listed in Appendix D. 

\section{\bf Amplitudes and branching ratios}
\subsection{Hadronization of currents}
The effective Lagrangians (the matrix elements) for photon and 
$Z$-boson part of the amplitudes for SL LFV lepton decays comprise vector 
and axial-vector currents, but box amplitude contains all possible quark 
currents permitted by Dirac algebra, that is scalar-, pseudoscalar-, 
vector-, axial-vector- and tensor-quark currents. To perform calculation 
of charged-lepton SL LFV decays rates these currents have be converted 
into meson currents comprising mesons that appear in the possible final 
products of the charged-lepton SL LFV decays we study here. The hadronization 
procedure we use here is not exact in the sense that we do not include 
the sea-quark and gluon content of the meson fields, but it is precise 
enough to give much better than order of magnitude decay rates of 
the processes considered here. The quark content of the meson states is 
given in Appendix E. The hadronization of axial-vector current is achieved through 
PCAC (see e.g. (\cite{Marshak69,Sakurai69,Fubini73}); for normalization 
of pseudoscalar coupling constants used here and for further details 
see \cite{Ilakovac:1995km}). The hadronization of the vector current is achieved 
using vector-meson-dominance assumption (see (\cite{Sakurai69,Fubini73}); 
for normalization of the vector meson-decay constant and details see 
\cite{Zielinski:1986mg,Ilakovac:1995km,Ilakovac:1995wc}). Hadronization of scalar currents is 
achieved by comparing the quark sector of the SM Lagrangian and the corresponding 
effective meson Lagrangian \cite{Bardeen:1986uz} 
(for applications in content of LFV and details see \cite{Ilakovac:1995wc}). 
Hadronization of the pseudoscalar current is obtained by
the same procedure as for the scalar current. 
The results obtained by using this procedure is equal to the result obtained 
by using equation of motion for current-quark masses (see e.g. \cite{Gasser:1983yg}) 
and results for hadronization of axial-vector current up to the difference 
of up and down quark masses or up to the difference of pseudoscalar decay 
constants. The hadronization of the tensor-quark currents is obtained by 
comparing the derivative of tensor-quark current with vector-quark current 
and using the equations of motion for current quark masses. The difference 
between terms, one containing the derivative of the incoming quark field and the 
other containing derivative of the outgoing quark field, have been neglected. 
The error expected from this approximation is proportional to the amount 
of breaking of $SU(3)_{\rm flavour}$ symmetry. The tensor currents are proportional 
to the current quark masses, and therefore give smaller contribution 
than the other quark currents. Therefore, the error introduced by 
this approximation in the total SL LFV amplitude is small.

Here we summarize the basic quantities needed to describe the hadronization 
of quark currents. 

\noindent
1. The pseudoscalar meson decay constants \cite{Eidelman:2004wy}
($f_P$, $P=\pi^0,\ \eta,\ \eta',\ K^0, \bar{K}^0$);
\\
2. The constants $\gamma_V$ \cite{Zielinski:1986mg}
($V= \rho^0,\ \phi,\ \omega,\ K^{*0}, \bar{K}^{*0}$) 
defining the vector meson decay constants ($f_V\sim m_V^2/\gamma_V$); 
\\
3. The mixing angles $\theta_P$ and $\theta_V$ \cite{Eidelman:2004wy} defining the physical 
meson-nonet states in terms of the unphysical singlet  and octet meson states;
\\
4. The parameter $r$ \cite{Bardeen:1986uz} ($m_u$, $m_d$ and $m_s$ are current quark masses), 
\bea
&&r\ =\ \frac{2m_{\pi^+}^2}{m_u+m_d}
 \ =\ \frac{2m_{K^0}^2}{m_d+m_s}
 \ =\ \frac{2m_{K^+}^2}{m_u+m_s},
\eea
that appears in the hadronization procedure for scalar and pseudoscalar 
currents. 

Having the identification of the quark currents with the corresponding 
meson currents, achieved by the above hadronization procedure, one can 
write down the effective Lagrangian as a sum of terms with an incoming 
lepton field $\ell_i$ and outgoing lepton field $\ell_j$ and pseudoscalar 
meson ($P$) or vector meson ($V$) field(s). This Lagrangian directly 
gives the amplitudes for the $\ell_i \to \ell_j P(V)$ processes. 
Amplitudes with a pseudoscalar meson in the final state contributions 
come from the pseudoscalar and axial-vector coupling part of the effective 
Lagrangian, while the amplitudes with vector mesons have vector and tensor 
coupling contributions. Only the scalar coupling gives no contribution to 
the one-meson processes in the final state, $\ell_i \to \ell_j P(V)$. 
They contribute only to the processes with two pseudoscalar mesons 
in the final state, $\ell_i \to \ell_j P_1P_2$. 

\subsection{Vector-meson--pseudoscalar-meson interactions}
The processes with two pseudoscalar mesons in the final state are 
generated by the scalar-quark-current part of the effective Lagrangian, 
and vector-quark- and tensor-quark-current of the effective Lagrangian. 
The scalar-quark-current part of the effective Lagrangian produces two 
pseudoscalar fields directly. The vector-quark- and tensor-quark-current 
parts produce a resonant vector meson state (V), which decays into 
two pseudoscalar mesons (P). The $VPP$ interactions necessary for 
description of the $VPP$ interactions appearing in the charged-lepton 
SL LFV decays are described by the part of the meson Lagrangian containing 
these $VPP$ vertices \cite{Ilakovac:1995wc}
\bea
\label{LVPP}
{\cal L}_{VPP} &=&
-\frac{ig_{\rho\pi\pi}}{2}
\Big\{\rho^{0,\mu}\big(
2\pi^+\hspace{-4pt}\stackrel{\leftrightarrow}{\partial}_\mu\hspace{-2pt}\pi^-
+K^+\hspace{-4pt}\stackrel{\leftrightarrow}{\partial}_\mu
     \hspace{-2pt}K^-
-K^0\hspace{-4pt}\stackrel{\leftrightarrow}{\partial}_\mu
     \hspace{-2pt}\bar{K}^0\big)
\nonumber\\
&+&\sqrt{3}s_V\omega^\mu\big(
  K^+\hspace{-4pt}\stackrel{\leftrightarrow}{\partial}_\mu
     \hspace{-2pt}K^-
 +K^0\hspace{-4pt}\stackrel{\leftrightarrow}{\partial}_\mu
     \hspace{-2pt}\bar{K}^0\big)
+\sqrt{3}c_V\phi^\mu\big(
  K^+\hspace{-4pt}\stackrel{\leftrightarrow}{\partial}_\mu
     \hspace{-2pt}K^-
 +K^0\hspace{-4pt}\stackrel{\leftrightarrow}{\partial}_\mu
     \hspace{-2pt}\bar{K}^0\big)
\nonumber\\
&+&K^{0*,\mu}\Big(
-\sqrt{2}
  \pi^+\hspace{-4pt}\stackrel{\leftrightarrow}{\partial}_\mu
     \hspace{-2pt}K^-
 +\pi^0\hspace{-4pt}\stackrel{\leftrightarrow}{\partial}_\mu
     \hspace{-2pt}\bar{K}^0
 +\sqrt{3}c_P
  \bar{K}^0\hspace{-4pt}\stackrel{\leftrightarrow}{\partial}_\mu
     \hspace{-2pt}\eta
 +\sqrt{3}s_P
  \bar{K}^0\hspace{-4pt}\stackrel{\leftrightarrow}{\partial}_\mu
     \hspace{-2pt}\eta'\Big)
\nonumber\\
&+&\bar{K}^{0*,\mu}\Big(
  \sqrt{2}
  \pi^-\hspace{-4pt}\stackrel{\leftrightarrow}{\partial}_\mu
     \hspace{-2pt}K^+
 -\pi^0\hspace{-4pt}\stackrel{\leftrightarrow}{\partial}_\mu
     \hspace{-2pt}K^0
 -\sqrt{3}c_P
  K^0\hspace{-4pt}\stackrel{\leftrightarrow}{\partial}_\mu
     \hspace{-2pt}\eta
 -\sqrt{3}s_P
  K^0\hspace{-4pt}\stackrel{\leftrightarrow}{\partial}_\mu
     \hspace{-2pt}\eta'\Big)
\Big\}+\dots.
\eea
This Lagrangian is a part of the nonlinear $(U(3)_L\times U(3)_R)/U(3)_V$ 
symmetric sigma-model Lagrangian. $U(3)_V$ symmetry corresponds to 
the vector mesons in the linear realization of the gauge equivalent 
$(U(3)_L\times U(3)_R)_{\rm global}\times U(3)_V$ linear sigma-model 
\cite{Bando:1987br,Bando:1985rf}. 
One can include the $(U(3)_L\times U(3)_R)/U(3)_V$ 
breaking terms, too \cite{Bando:1985rf}. That was applied to SL LFV tau-lepton 
decays in Ref. \cite{Ilakovac:1995wc}, but for the estimates of 
the charged-lepton SL LFV decays it is unnecessary complication, 
and we will not consider it here. 

From Eq. (\ref{LVPP}) one can read of the $c_{VP_1P_2}$ couplings in 
terms of $g_{\rho\pi\pi}$ coupling. For instance, 
$c_{\rho^0K^+K^-}=\frac{1}{2}g_{\rho\pi\pi}$. 

When the amplitudes with vector-meson resonance(s) are formed, the square 
of the vector meson mass in the $m_V^2/\gamma_V$, appearing 
in every vector meson decay constant, has to be replaced with 
$(m_V^2-im_V\Gamma_V)/\gamma_V$, where $\Gamma_V$ 
is the decay width of the vector meson \cite{Ilakovac:1995wc}. 

\subsection{Charged-lepton SL LFV with one meson in the final state}
Now we can write down all amplitudes we are interested in. 
(Notice that the axial-vector mesons ($A$) are not included. They decay 
into three pseudoscalar mesons, and therefore the amplitudes are much 
more complicated. New vertices with $A-V-P$ couplings should be included, 
kinematics is much more involved {\it etc.} The scalar mesons are also not 
included. They also lead to complications.). They are given as follows. 
\bea
\label{MliljV}
i{\cal M}^{\ell_i\to \ell_j V} &=&
i \bar{u}_{\ell_j} \left[
\left(\gamma_\mu-\frac{q_\mu\gamma\cdot q}{q^2}\right) P_L 
{\cal P}_{1\gamma L}^{ijV}
+\left(\gamma_\mu-\frac{q_\mu\gamma\cdot q}{q^2}\right) P_R 
{\cal P}_{1\gamma R}^{ijV}
\right.
\nonumber\\ 
&+&\frac{i\sigma_{\mu\nu} P_L q^\nu}{q^2} {\cal P}_{2\gamma L}^{ijV}
+\frac{i\sigma_{\mu\nu} P_R q^\nu}{q^2} {\cal P}_{2\gamma R}^{ijV}
\nonumber\\
&+&\gamma_\mu P_L \left( {\cal P}_{ZL}^{ijV}+{\cal B}_{1L}^{ijV} \right)
+\gamma_\mu P_R \left( {\cal P}_{ZR}^{ijV}+{\cal B}_{1R}^{ijV} \right)
\nonumber\\
&+&\left.
i\sigma_{\mu\nu} P_L q^\nu {\cal B}_{2L}^{ijV}
+i\sigma_{\mu\nu} P_R q^\nu {\cal B}_{2R}^{ijV}
\right] u_{\ell_i} \varepsilon^{\mu\dagger}_V. 
\eea
\bea
\label{MliljP}
i{\cal M}^{\ell_i \to \ell_j P} &=&
i \bar{u}_{\ell_j} \left\{
\left[ \gamma_\mu P_L \left( {\cal P}_{ZL}^{ijP}
+{\cal B}_{1L}^{ijP} \right)
+\gamma_\mu P_R \left( {\cal P}_{ZR}^{ijP}
+{\cal B}_{1R}^{ijP} \right) \right] q^\mu
\right.
\nonumber\\
&+&\left.
\left[ P_L {\cal B}_{2L}^{ijP} + P_R {\cal B}_{2R}^{ijP}\right]
\right\} u_{\ell_i}. 
\eea
The form factors in Eqs. (\ref{MliljV}) and (\ref{MliljP}) 
are defined as follows:
\bea
%
{\cal P}_{a\gamma\: L,R}^{ijV}
\label{VgFF}
&=& 
-e^2 
{\cal P}^{L,R}_{a\gamma} 
\frac{m_{V}}{\sqrt{2}\gamma_{V}}
k^{V}_\gamma,
\qquad (a=1,2),\ \ (V=\rho^0,\phi,\omega)
\\ 
%
{\cal P}_{ZL,R}^{ijV}
\label{VZFF}
&=&
\frac{g_Z^2}{m_Z^2c_W^2}
{\cal P}^{L,R}_{Z}
\frac{m_V^2}{\sqrt{2}\gamma_V}
k^{V}_Z,     
\qquad\qquad (V=\rho^0,\phi,\omega)
\\
%
{\cal B}_{1L,R}^{ijV}
\label{VbFF}
&=&
\frac{m_V^2}{\sqrt{2}\gamma_V}\frac{1}{2}
\left[
k_{\bar{u}u}^V
\left({\cal{B}}_{1\bar{u}u}^{L,R}
+{\cal{B}}_{2\bar{u}u}^{L,R}\right)
+
k_{\bar{d}d}^V
\left({\cal{B}}_{1\bar{d}d}^{L,R}
+{\cal{B}}_{2\bar{d}d}^{L,R}\right)
\right.
\nonumber\\
&+&
k_{\bar{s}s}^V
\left({\cal{B}}_{1\bar{s}s}^{L,R}
+{\cal{B}}_{2\bar{s}s}^{L,R}\right)
+
k_{\bar{d}s}^V
\left({\cal{B}}_{1\bar{d}s}^{L,R}
+{\cal{B}}_{2\bar{d}s}^{L,R}\right)
\nonumber\\&&
\left.
+
k_{\bar{s}d}^V
\left({\cal{B}}_{1\bar{s}d}^{L,R}
+{\cal{B}}_{2\bar{s}d}^{L,R}\right)
\right],\qquad\qquad
(V=\rho^0,\phi,\omega,K^{*0},\overline{K^{*0}}),
\\
{\cal B}_{2L,R}^{ijV}
&=&
\frac{-2\sqrt{2}}{\gamma_V}
\left[
k_{\bar{d}s}^V
(m_d-m_s) 
{\cal{B}}_{4\bar{d}s}^{L,R}
\right.
\nonumber\\&& 
+
\left.
k_{\bar{s}d}^V
(m_s-m_d)
{\cal{B}}_{4\bar{s}d}^{L,R}
\right],\qquad\qquad (V=K^{*0},\overline{K^{*0}}),
\\
%
{\cal P}_{ZL,R}^{ijP} &=&
\label{PZFF}
\frac{g^2}{m_Z^2c_W^2} {\cal P}^{L,R}_{Z} (-\sqrt{2}f_P) k_Z^P, 
\qquad\qquad (P=\pi^0,\eta,\eta')
\\
%
{\cal B}_{1L,R}^{ijP}
\label{Pb1FF}
&=&
s_{L,R} (\sqrt{2}f_P) \frac{1}{2}
\left[
k_{\bar{u}u}^P
\left(-{\cal{B}}_{1\bar{u}u}^{L,R}
+{\cal{B}}_{2\bar{u}u}^{L,R}\right)
+
k_{\bar{d}d}^P
\left(-{\cal{B}}_{1\bar{d}d}^{L,R}
+{\cal{B}}_{2\bar{d}d}^{L,R}\right)
\right.
\nonumber\\
&&
+
k_{\bar{s}s}^P
\left({-\cal{B}}_{1\bar{s}s}^{L,R}
+{\cal{B}}_{2\bar{s}s}^{L,R}\right)
+
k_{\bar{d}s}^P
\left(-{\cal{B}}_{1\bar{d}s}^{L,R}
+{\cal{B}}_{2\bar{d}s}^{L,R}\right)
\nonumber\\&&
\left.
+
k_{\bar{s}d}^P
\left(-{\cal{B}}_{1\bar{s}d}^{L,R}
+{\cal{B}}_{2\bar{s}d}^{L,R}\right)
\right], 
\qquad\qquad
(P=\pi^0,\eta,\eta',K^{0},\overline{K^{0}}),
\\
%
{\cal B}_{2L,R}^{ijP}
\label{Pb2FF}
&=&
s_{L,R} (-\frac{ir}{2})(\sqrt{2}f_P) \frac{1}{2}
\left[
k_{\bar{u}u}^P
\left(-{\cal{B}}_{3\bar{u}u}^{L,R}
+\overline{{\cal{B}}}_{3\bar{u}u}^{L,R}\right)
+
k_{\bar{d}d}^P
\left(-{\cal{B}}_{3\bar{d}d}^{L,R}
+\overline{{\cal{B}}}_{3\bar{d}d}^{L,R}\right)
\right.
\nonumber\\
&&
+
k_{\bar{s}s}^P
\left({-\cal{B}}_{3\bar{s}s}^{L,R}
+\overline{{\cal{B}}}_{3\bar{s}s}^{L,R}\right)
+
k_{\bar{d}s}^P
\left(-{\cal{B}}_{3\bar{d}s}^{L,R}
+\overline{{\cal{B}}}_{3\bar{d}s}^{L,R}\right)
\nonumber\\&&
\left.
+
k_{\bar{s}d}^P
\left(-{\cal{B}}_{3\bar{s}d}^{L,R}
+\overline{{\cal{B}}}_{3\bar{s}d}^{L,R}\right)
\right],
\qquad\qquad
(P=\pi^0,\eta,\eta',K^{0},\overline{K^{0}})\;.
\eea
In Eqs. (\ref{Pb1FF}) and (\ref{Pb2FF}) 
$s_L=1$ and $s_R=-1$. The constants 
$k^V_\gamma$, $k^V_Z$, $k^P_Z$, and $k^V_{\bar{q}_aq_b}$ 
are defined in Appendix E.

A branching ratio for the processes $\ell_i\to\ell_j P$ 
with unpolarized initial and final particles reads
\bea
B(\ell_i \to \ell_j P) &=&
\frac{1}{8\pi} \frac{1}{m_i^2} \frac{1}{\Gamma_{l_i}}
\frac{\lambda^{\frac{1}{2}}(m_i^2,m_j^2,m_P^2)}{2m_i}
\nonumber\\
&\times& \Bigg[
\left(
|{\cal P}_{ZL}^{ijP}+{\cal B}_{1L}^{ijP}|^2
+ |{\cal P}_{ZR}^{ijP}+{\cal B}_{1R}^{ijP}|^2
\right)
i_{P1}
\nonumber\\
&+&
\left(
|{\cal B}_{2L}^{ijP}|^2 + |{\cal B}_{2R}^{ijP}|^2
\right)
i_{P2}
\nonumber\\
\nonumber\\
&+&
\left(
({\cal P}_{ZL}^{ijP}+{\cal B}_{1L}^{ijP})
({\cal P}_{ZR}^{ijP}+{\cal B}_{1R}^{ijP})^*
+ c.c.
\right)
m_jm_i i_{P3}
\nonumber\\
&+&
\left(
({\cal P}_{ZL}^{ijP}+{\cal B}_{1L}^{ijP})({\cal B}_{2L}^{ijP})^*
+ ({\cal P}_{ZR}^{ijP}+{\cal B}_{1R}^{ijP})({\cal B}_{2R}^{ijP})^*
+ c.c.
\right)
m_j i_{P4}
\nonumber\\
&+&
\left(
({\cal P}_{ZL}^{ijP}+{\cal B}_{1L}^{ijP})({\cal B}_{2R}^{ijP})^*
+ ({\cal P}_{ZR}^{ijP}+{\cal B}_{1R}^{ijP})({\cal B}_{2L}^{ijP})^*
+ c.c.
\right)
m_i i_{P5}
\nonumber\\
&+&
\left(
({\cal B}_{2L}^{ijP})({\cal B}_{2R}^{ijP})^* + c.c.
\right)
m_i m_j
\Bigg],
\eea
where $\Gamma_{l_i}$ is a total decay rate of the lepton $l_i$ and
\bea
i_{P1} &=& \frac{1}{2} \left( (m_i^2-m_j^2)^2-(m_i^2+m_j^2)m_P^2 \right),
\nonumber\\
i_{P2} &=& \frac{1}{2}(m_i^2+m_j^2-m_P^2),
\nonumber\\
i_{P3} &=& m_P^2,
\nonumber\\
i_{P4} &=& \frac{1}{2}(m_i^2+m_P^2-m_j^2),
\nonumber\\
i_{P5} &=& \frac{1}{2}(m_i^2-m_P^2-m_j^2),
\eea
and
\be
\lambda(x,y,z)\ =\ x^2+y^2+z^2-2xy-2xz-2yz.
\ee

A branching ratio for the processes $\ell_i \to \ell_j V$
with unpolarized initial and final particles reads.
\bea
B(\ell_i \to \ell_j V) &=&
\frac{1}{8\pi} \frac{1}{\Gamma_{\ell_i}}\frac{1}{m_i^2}
\frac{\lambda^{\frac{1}{2}}(m_i^2,m_j^2,m_V^2)}{2m_i}
\nonumber\\
&\times& \Bigg[
\left(
|{\cal P}_{1\gamma L}^{ijV}+{\cal P}_{ZL}^{ijV}
+{\cal B}_{1L}^{ijV}|^2
+ |{\cal P}_{1\gamma R}^{ijV}+{\cal P}_{ZR}^{ijV}
+{\cal B}_{1R}^{ijV}|^2
\right)
i_{V1}
\nonumber\\
&+&
\left(
\left| \frac{{\cal P}_{2\gamma L}^{ijV}}{m_V^2}
+ {\cal B}_{2L}^{ijV} \right |^2
+ \left|\frac{{\cal P}_{2\gamma R}^{ijV}}{m_V^2}
+ {\cal B}_{2R}^{ijV} \right |^2
\right)
i_{V2}
\nonumber\\
&+&
\left(
({\cal P}_{1\gamma L}^{ijV}+{\cal P}_{ZL}^{ijV}
+{\cal B}_{1L}^{ijV})
({\cal P}_{1\gamma R}^{ijV}+{\cal P}_{ZR}^{ijV}
+{\cal B}_{1R}^{ijV})^*
+c.c.
\right)
(-m_im_j)
\nonumber\\
&+&
\left(
\left( \frac{{\cal P}_{2\gamma L}^{ijV}}{m_V^2}
+ {\cal B}_{2L}^{ijV} \right)
\left( \frac{{\cal P}_{2\gamma R}^{ijV}}{m_V^2}
+ {\cal B}_{2R}^{ijV} \right)^*
+c.c.
\right)
(-m_V^2m_im_j)
\nonumber\\
&+&
\Bigg(
({\cal P}_{1\gamma L}^{ijV}+{\cal P}_{ZL}^{ijV}
+{\cal B}_{1L}^{ijV})
\left( \frac{{\cal P}_{2\gamma L}^{ijV}}{m_V^2}
+ {\cal B}_{2L}^{ijV} \right)^*
\nonumber\\
&+&
({\cal P}_{1\gamma R}^{ijV}+{\cal P}_{ZR}^{ijV}
+{\cal B}_{1R}^{ijV})
\left( \frac{{\cal P}_{2\gamma R}^{ijV}}{m_V^2}
+ {\cal B}_{2R}^{ijV} \right)^*
+c.c.
\Bigg)
(m_j i_{V3})
\nonumber\\
&+&
\Bigg(
({\cal P}_{1\gamma L}^{ijV}+{\cal P}_{ZL}^{ijV}
+{\cal B}_{1L}^{ijV})
\left( \frac{{\cal P}_{2\gamma R}^{ijV}}{m_V^2}
+ {\cal B}_{2R}^{ijV} \right)^*
\nonumber\\
&+&
({\cal P}_{1\gamma R}^{ijV}+{\cal P}_{ZR}^{ijV}
+{\cal B}_{1R}^{ijV})
\left( \frac{{\cal P}_{2\gamma L}^{ijV}}{m_V^2}
+ {\cal B}_{2L}^{ijV} \right)^*
+c.c.
\Bigg)
(m_i i_{V4})
\Bigg],
\eea
where
\bea
i_{V1} &=&
\frac{1}{2m_V^2} \left[ m_V^2 (m_i^2+m_j^2) + (m_i^2-m_j^2)^2 - 2 m_V^4 \right],
\nonumber\\
i_{V2} &=&
(m_i^2-m_j^2)^2 - \frac{1}{2} m_V^2 (m_i^2+m_j^2) -\frac{1}{2} m_V^4,
\nonumber\\
i_{V3} &=& \frac{1}{2}\left( m_i^2-m_j^2+m_V^2 \right),
\nonumber\\
i_{V4} &=& \frac{1}{2}\left( m_i^2-m_j^2-m_V^2 \right).
\eea

\subsection{SL LFV decays of a lepton with two pseudoscalar mesons 
in the final state}
Amplitude for a general $\ell_i \to \ell_j P_1 P_2$ decay rate is 
a sum of scalar-coupling contribution and resonance contributions 
(coming from vector- and tensor-coupling contributions)
\be
i{\cal M}^{\ell_i \to \ell_j P_1 P_2} \ =\
i{\cal M}_{res}^{\ell_i \to \ell_j P_1 P_2} 
+ i{\cal M}_{1}^{\ell_i \to \ell_j P_1 P_2},
\ee
where
\bea
{\cal M}_{res}^{\ell_i \to \ell_j P_1 P_2}
\label{Mres}
&=&
i \bar{u}_{\ell_j}
\Bigg[
D_{1L}^{P_1P_2} \left( (\Slash{p}_2-\Slash{p}_1)
-\frac{m_2^2-m_1^2}{q^2}\Slash{q} 
\right) P_L
\nonumber\\
&+& D_{1R}^{P_1P_2} \left( (\Slash{p}_2-\Slash{p}_1)
-\frac{m_2^2-m_1^2}{q^2}\Slash{q} 
\right) P_R
\nonumber\\
&+& E_{1L}^{P_1P_2} i\sigma_{\mu\nu} P_L (p_1-p_2)^\mu q^\nu
+ E_{1R}^{P_1P_2} i\sigma_{\mu\nu} P_R (p_1-p_2)^\mu q^\nu\Bigg] u_{\ell_i},
\\
i{\cal M}_{1}^{\ell_i \to \ell_j P_1 P_2} &=&
i \bar{u}_{\ell_j} 
\left[
P_L A_{1P_1P_2}^L + P_R A_{1P_1P_2}^R
\right] u_{\ell_i}.
\eea
In Eq. (\ref{Mres}) $D_{1L,R}^{P_1P_2}$ and $E_{1L,R}^{P_1P_2}$ 
are form factors built from the trilinear $c_{VP_1P_2}$ couplings 
(defined by the Lagrangian (\ref{LVPP})), normalized vector-meson 
propagators, 
\be
\frac{m_V^2-im_V \Gamma_V}{q^2-m_V^2+im_V \Gamma_V}
\ee
and form factors for $\ell_i\to \ell_j V$ processes divided by the 
the mass of the resonant vector meson, e.g. 
\bea
\tilde{{\cal P}}_{1\gamma L,R}^{ijV} &=&
\frac{{\cal P}_{1\gamma L,R}^{ijV}}{m_V^2}.
\eea
The expressions for the $D_{1L,R}^{P_1P_2}$ and $E_{1L,R}^{P_1P_2}$
form factors are
\bea
D_{1L,R}^{P_1 P_2} &=&
\sum_V 
\left(
\tilde{{\cal{P}}}_{1\gamma L,R}^{ijV} 
+ \tilde{{\cal{P}}}_{ZL,R}^{ijV} 
+ \tilde{{\cal{B}}}_{1L,R}^{ijV} \right)
\frac{m_V^2-im_V \Gamma_V}{q^2-m_V^2+im_V \Gamma_V} c_{VP_1P_2}, 
\\
E_{1L,R}^{P_1P_2} &=&
\sum_V
\left(
\frac{\tilde{{\cal P}}_{2\gamma L,R}^{ijV}}{q^2} 
+ \tilde{{\cal B}}_{2L,R}^{ijV}
\right)
\frac{m_V^2-im_V\Gamma_V}{q^2-m_V^2+im_V\Gamma_V} c_{VP_1P_2}. 
\eea
The sum goes over neutral vector mesons only 
$(V=\rho^0,\phi,\omega,K^{*0},\overline{K^{*0}})$. The coefficients of 
the non-resonant part of the amplitude, the constants $A_{1P_1P_2}$, 
are defined as the coefficients of the $P_1P_2$ product of fields contained 
in the matrix-valued operator 
\bea
\frac{r}{4}
\sum_{q_a,q_b=u,d,s}{(\Pi^2)_{\bar{q}_bq_a}
(\cal{B}}_{3\bar{q}_aq_b}^{L,R}
+\overline{{\cal{B}}}_{3\bar{q}_aq_b}^{L,R})
\eea
where $\Pi$ is a matrix of pseudoscalar fields
\bea
\Pi &=& 
\left(
\begin{array}{ccc}
\pi^0 +
\frac{1}{\sqrt{3}}\eta_8+
\frac{\sqrt{2}}{\sqrt{3}}\eta_1 
&
\sqrt{2}\pi^+
&
\sqrt{2} K^+
\\
\sqrt{2}\pi^-
&
-\pi^0 +
\frac{1}{\sqrt{3}}\eta_8+
\frac{\sqrt{2}}{\sqrt{3}}\eta_1
&
\sqrt{2} K^0
\\
\sqrt{2} K^-
&
\sqrt{2} \bar{K}^0
&
-\frac{2}{\sqrt{3}}\eta_8+
\frac{\sqrt{2}}{\sqrt{3}}\eta_1
\end{array}
\right).
\eea
For example,
\bea
A_{1\pi^0\pi^0}^{L,R} &=&
\frac{r}{4}
\left[
\frac{1}{2}
\left(
{\cal{B}}_{3\bar{u}u}^{L,R}
+\overline{{\cal{B}}}_{3\bar{u}u}^{L,R}
+{\cal{B}}_{3\bar{d}d}^{L,R}
+\overline{{\cal{B}}}_{3\bar{d}d}^{L,R}
\right)
\right].
\eea
Having the amplitudes one can easily evaluate the branching fractions. 
We assume that incoming and outgoing particles are not polarised. 
\bea
B(\ell_i\to \ell_j P_1P_2) &=&
\frac{1}{(2\pi)^2} \frac{1}{\Gamma_{\ell_i}} \frac{1}{32m_i^3}
\int_{(m_1+m_2)^2}^{(m_i-m_j)^2} ds_{12}
\int_{s_{j2}^{min}}^{s_{j2}^{max}} ds_{j2}\
|{\cal M}^{\ell_i \to \ell_j P_1P_2}|^2
\nonumber\\
&=&
\frac{1}{(2\pi)^2}\frac{1}{32m_i^3}
\int_{(m_1+m_2)^2}^{(m_i-m_j)^2} ds_{12}
\nonumber\\
&\times&
\Bigg[
\left( |D_{1L}^{P_1P_2}|^2 + |D_{1R}^{P_1P_2}|^2 \right) \bar{I}_1
\nonumber\\
&+&
\left( |E_{1L}^{P_1P_2}|^2 + |E_{1R}^{P_1P_2}|^2 \right) \bar{I}_2
\nonumber\\
&+&
\left( |A_{1P_1P_2}^L|^2 + |A_{1P_1P_2}^R|^2 \right) \bar{I}_3
\nonumber\\
&+&
\left( (D_{1L}^{P_1P_2})(D_{1R}^{P_1P_2})^* +c.c. \right) \bar{I}_4
\nonumber\\
&+&
\left( (E_{1L}^{P_1P_2})(E_{1R}^{P_1P_2})^* +c.c. \right) \bar{I}_5
\nonumber\\
&+&
\left( (A_{1P_1P_2}^L)(A_{1P_1P_2}^R)^* + c.c. \right) \bar{I}_6
\nonumber\\
&+&
\left( (D_{1L}^{P_1P_2})(E_{1L}^{P_1P_2})^*
 + (D_{1R}^{P_1P_2})(E_{1R}^{P_1P_2})^* + c.c. \right) \bar{I}_7
\nonumber\\
&+&
\left( (D_{1L}^{P_1P_2})(E_{1R}^{P_1P_2})^*
 + (D_{1R}^{P_1P_2})(E_{1L}^{P_1P_2})^* + c.c. \right) \bar{I}_8
\nonumber\\
&+&
\left( (D_{1L}^{P_1P_2})(A_{1P_1P_2}^L)^*
 + (D_{1R}^{P_1P_2})(A_{1P_1P_2}^R)^* + c.c. \right) \bar{I}_9
\nonumber\\
&+& \left( (D_{1L}^{P_1P_2})(A_{1P_1P_2}^R)^*
 + (D_{1R}^{P_1P_2})(A_{1P_1P_2}^L)^* + c.c. \right) \bar{I}_{10}
\nonumber\\
&+&
\left( (E_{1L}^{P_1P_2})(A_{1P_1P_2}^L)^*
 + (E_{1R}^{P_1P_2})(A_{1P_1P_2}^R)^* + c.c. \right) \bar{I}_{11}
\Bigg].
\eea
The Mandelstam variables are defined as $s_{ab}=(p_a-p_b)^2$, e.g. 
$s_{12} = (p_1-p_2)^2$. The kinematical bounds on the Mandelstam 
variables $s_{j2}^{min}$ and $s_{j2}^{max}$ are well known \cite{Eidelman:2004wy}. 
The $\bar{I}$ integrals read, 
\bea
\bar{I}_1 &=& 2 \overline{s_{j2}^2} + 2 \overline{s_{j2}} (e_1+e_2)
 +(2e_1e_2-e_3e_4) \overline{1},
\nonumber\\
\bar{I}_2 &=& - 2 \overline{s_{j2}^2} (e_{10}) 
 + 2 \overline{s_{j2}} (e_5e_{10}+e_9e_{10}-e_6e_7-e_8e_7)
\nonumber\\
&+& 2 ((e_5e_6e_7+e_8e_9e_7-e_5e_9e_{10}-e_8e_6e_{11}) 
+ e_3(e_{11}e_{10}-e_7^2)) 
   \overline{1},
\nonumber\\
\bar{I}_3 &=& e_3 \overline{1},
\nonumber\\
\bar{I}_4 &=& m_im_j e_4 \overline{1},
\nonumber\\
\bar{I}_5 &=& m_im_j (e_{10}e_{11}-e_7^2) \overline{1},
\nonumber\\
\bar{I}_6 &=& m_im_j \overline{1},
\nonumber\\
\bar{I}_7 &=& m_j (-e_{12}e_6) \overline{1},
\nonumber\\ 
\bar{I}_8 &=& m_i (e_{12}e_8) \overline{1},
\nonumber\\
\bar{I}_9 &=& m_j \left( \overline{s_{j2}} + (e_2) \overline{1} \right),
\nonumber\\
\bar{I}_{10} &=& m_i \left( \overline{s_{j2}} + (e_1) \overline{1} \right),
\nonumber\\
\bar{I}_{11} &=& - \overline{s_{j2}} (e_8-e_6) 
+ (e_8e_9-e_5e_6) \overline{1},
\eea
where 
\be
\overline{s_{j2}^n}\ =\
\int_{s_{j2}^{min}}^{s_{j2}^{max}} ds_{j2}\ s_{j2}^n. 
\ee
The quantities $e_i$ read 
\bea
e_1 &=& - e_5 - \frac{m_2-m_1}{s_{12}} e_8,
\nonumber\\
e_2 &=& - e_9 - \frac{m_2-m_1}{s_{12}} e_6,
\nonumber\\
e_3 &=& \frac{1}{2}(m_i^2+m_j^2-s_{12}),
\nonumber\\
e_4 &=& e_{11}-\frac{(m_2^2-m_1^2)^2}{s_{12}},
\nonumber\\
e_5 &=& m_2^2  + \frac{1}{2} (m_i^2+m_j^2-s_{12}),
\nonumber\\
e_6 &=& \frac{1}{2} (m_i^2-m_j^2+s_{12}),
\nonumber\\
e_7 &=& m_1^2-m_2^2,
\nonumber\\
e_8 &=& \frac{1}{2} (m_i^2-m_j^2-s_{12}),
\nonumber\\
e_9 &=& m_1^2 + \frac{1}{2} (m_i^2+m_j^2-s_{12}),
\nonumber\\
e_{10} &=& s_{12},
\nonumber\\
e_{11} &=& 2m_1^2 + 2m_2^2 - s_{12},
\nonumber\\
e_{12} &=& -e_4.
\eea

\section{Minimal SO(10) model and its predictions}
Now we are ready to estimate the SL LFV processes. In order to perform 
a concrete evaluation for the SL LFV processes, we need an information 
on the Yukawa couplings. In this paper, we make use of the minimal 
$SO(10)$ model, as an example which gives a precise information for 
the neutrino-Dirac-Yukawa couplings.  We begin with an overreview of the minimal 
SUSY $SO(10)$ model proposed in \cite{Babu} and recently analysed in detail 
in Ref. \cite{Fukuyama:2002ch, Goh:2003sy, Bajc:2002iw, Dutta:2004wv, Matsuda:2004bq}. 
Even when we concentrate our discussion on the issue how to 
reproduce the realistic fermion mass matrices in the $SO(10)$ model, 
there are lots of possibilities for introduction of Higgs multiplets. 
The minimal supersymmetric $SO(10)$ model 
is the one where only  one {\bf 10} and one $\overline{\bf 126}$ Higgs 
multiplet have Yukawa couplings (superpotential) with {\bf 16} matter 
multiplets. Therefore, the quark and lepton mass matrices can be 
described as
\begin{eqnarray}
 M_u &=& c_{10} M_{10} + c_{126} M_{126}, 
 \nonumber \\
 M_d &=& M_{10} + M_{126},   
 \nonumber \\
 M_D &=& c_{10} M_{10} - 3 c_{126} M_{126}, 
 \nonumber \\
 M_e &=& M_{10} - 3 M_{126}, 
 \nonumber \\ 
 M_R &=& c_R M_{126}, 
\label{massmatrix}
\end{eqnarray} 
where $M_u$, $M_d$, $M_D$, $M_e$ and $M_R$ denote up-type quark, 
down-type quark, neutrino Dirac, charged-lepton and right-handed 
neutrino Majorana mass matrices, respectively. Note that all the quark 
and lepton mass matrices are characterized by only two basic mass matrices, 
$M_{10}$ and $M_{126}$, and three complex coefficients $c_{10}$, $c_{126}$ 
and $c_R$. 

The mass matrix formulas in Eq.~(\ref{massmatrix}) lead to the GUT relation 
among the quark and lepton mass matrices, 
\begin{eqnarray}
M_e = c_d \left( M_d + \kappa  M_u \right) \; , 
\label{GUTrelation} 
\end{eqnarray} 
where 
\begin{eqnarray}
c_d &=& - \frac{3 c_{10} + c_{126}}{c_{10}-c_{126}}, 
\\
\kappa &=& - \frac{4}{3 c_{10} + c_{126}}. 
\end{eqnarray} 
Without loss of generality, we can start with the basis where $M_u$ is 
real and diagonal, $M_u = D_u$. Since $M_d$ is the symmetric matrix, 
it can be described as 
$M_d = V_{\mathrm{CKM}}^* \,D_d \,V_{\mathrm{CKM}}^\dagger$ 
by using the CKM matrix $V_{\mathrm{CKM}}$ and the real diagonal 
mass matrix $D_d$.
\footnote{
In general, $M_d = U^* \,D_d \, U^\dagger$ 
by using a general unitary matrix 
$U=e^{i \alpha} e^{i \beta T_3} e^{i \gamma T_8} V_{\mathrm{CKM}} 
   e^{i \beta^\prime T_3} e^{i \gamma^\prime T_8}$. 
We omit the diagonal phases to keep the number of the free parameters 
in the model as small as possible. } 
Considering the basis-independent quantities, 
$\mathrm{tr} [M_e^\dagger M_e ]$, $\mathrm{tr} [(M_e^\dagger M_e)^2 ]$ 
and $\mathrm{det} [M_e^\dagger M_e ]$, and eliminating $|c_d|$, 
we obtain two independent equations,  
\begin{eqnarray}
\left(
\frac{\mathrm{tr} [\widetilde{M_e}^\dagger \widetilde{M_e} ]}
{m_e^2 + m_{\mu}^2 + m_{\tau}^2} \right)^2
&=& 
\frac{\mathrm{tr} [( \widetilde{M_e}^\dagger \widetilde{M_e} )^2 ]}
{m_e^4 + m_{\mu}^4 + m_{\tau}^4},
\label{cond1} \\ 
\left( \frac{\mathrm{tr} [\widetilde{M_e}^\dagger \widetilde{M_e} ]}
{m_e^2 + m_{\mu}^2 + m_{\tau}^2} \right)^3
&=&
\frac{\mathrm{det} [\widetilde{M_e}^\dagger \widetilde{M_e} ]}
{m_e^2 \; m_\mu^2 \; m_\tau^2},
\label{cond2} 
\end{eqnarray}
where $\widetilde{M_e} \equiv 
V_{\mathrm{CKM}}^* \, D_d \, V_{\mathrm{CKM}}^\dagger + \kappa D_u$. 
With input data of six quark masses, three angles and one CP-phase 
in the CKM matrix and three charged-lepton masses, we can solve 
the above equations and determine $\kappa$ and $|c_d|$, but one parameter, 
the phase of $c_d$, is left undetermined \cite{Fukuyama:2002ch}. 
The original basic mass matrices, $M_{10}$ and $M_{126}$, are described by 
\begin{eqnarray}
M_{10} 
&=& 
\frac{3+ |c_d| e^{i \sigma}}{4} 
V_{\mathrm{CKM}}^* \, D_d \, V_{\mathrm{CKM}}^\dagger
+ \frac{|c_d| e^{i \sigma} \kappa}{4} D_u, 
\label{M10}  \\ 
M_{126} &=& 
 \frac{1- |c_d| e^{i \sigma}}{4} 
 V_{\mathrm{CKM}}^* \, D_d \, V_{\mathrm{CKM}}^\dagger
 -\frac{|c_d| e^{i \sigma} \kappa}{4} D_u, 
 \label{M126} 
\end{eqnarray} 
as the functions of $\sigma$, the phase of $c_d$, with the solutions, 
 $|c_d|$ and $\kappa$, determined by the GUT relation. 

Now let us solve the GUT relation and determine $|c_d|$ and $\kappa$. 
Since the GUT relation of Eq.~(\ref{GUTrelation}) is valid only 
at the GUT scale, we first evolve by renormalization group equations (RGE's) 
the data from the weak scale 
to the corresponding quantities at the GUT scale for a given $\tan \beta$ 
Then we and use them as 
input data at the GUT scale. Note that it is non-trivial to find 
a solution of the GUT relation, since the number of free parameters 
(fourteen) is almost the same as the number of inputs (thirteen). 
The solution of the GUT relation exists only if we take appropriate 
input parameters. By using the experimental data at the $M_Z$ scale 
\cite{Fusaoka:1998vc}, we get the following absolute values for charged fermion 
masses (in units of GeV) and the CKM matrix at the GUT scale, 
$M_G$, with $\tan \beta = 45$,
$m_s = 0.072$ and $\delta = 1.518$: 
\begin{eqnarray}
& & m_u = 0.00103 \; , \; \;   m_c=0.299 \; , 
\; \;  m_t=133 \nonumber \\
& & m_d=0.00170 \; , \; \;  m_s =0.0263   \; , 
\; \;   m_b=1.55  \nonumber \\ 
& & m_e=0.000411 \; , \; \; m_\mu=0.0868 \; ,
\; \;  m_\tau=1.69  \nonumber 
\end{eqnarray} 
and 
\begin{eqnarray}
 V_{\mathrm{CKM}} = \left( 
 \begin{array}{ccc}
0.975 & 0.222 & 0.000146 - 0.00279 i \\
-0.222-0.000121 i & 0.974 + 0.000129 i & 0.0320 \\ 
0.00697 - 0.00272 i & -0.0312 - 0.000626 i & 0.999
   \end{array} 
  \right) \; ,  \nonumber
\end{eqnarray}
in the standard parameterization. The phases of the fermion masses are 
not determined by the diadonalization procedure. Here the masses are chosen 
be real. The signs of the fermion masses 
have been chosen to be equal, $-$ for $m_u$, $m_c$, $m_d$ and $m_s$, 
and $+$ for $m_t$ and $m_b$. So determined  masses at the GUT scale are used as 
input parameters in order to solve Eqs.~(\ref{cond1}) and (\ref{cond2}). 
As an example, here we show one of two solutions, 
\bea
\kappa &=&
0.0134 - 0.000791\; i \,,
\nonumber\\
|c_d| &=& 6.39  \, . 
\eea
Once the parameters, $|c_d|$ and $\kappa$, are determined, we can describe 
all the fermion mass matrices as a functions of $\sigma$ from the mass matrix 
formulas of Eqs.~(\ref{massmatrix}), (\ref{M10}) and (\ref{M126}). 
Interestingly, in the minimal $SO(10)$ model even light Majorana neutrino 
mass matrix, $M_{\nu}$, can be determined as a function of the phase $\sigma$ and 
$c_R$ through the seesaw mechanism $M_{\nu}= - M_D^T M_R^{-1} M_D$. 

Now we give an example of the neutrino-Dirac-Yukawa coupling matrix 
which can fit the GUT relation. In the basis where both 
the charged-lepton and right-handed Majorana neutrino mass matrix  
are diagonal with real and positive eigenvalues, the neutrino Dirac 
Yukawa coupling matrix at the GUT scale for fixed 
$\sigma=3.223\; \mathrm{[rad]}$ is found to be 
%
%
\be 
Y_{\nu} = 
\left( 
 \begin{array}{ccc}
 0.000310 + 0.00348 i & -0.000894 - 0.000249 i  &  0.0447   + 0.0531 i  \\ 
 0.00590  - 0.0103 i  & -0.0164   - 0.0427   i  &  0.308    + 0.116   i  \\ 
 0.00215   + 0.00126 i & 0.0558   - 0.0559 i  &  -0.381  + 0.604 i 
 \end{array}   \right) \; .  
\label{Ynu}
\ee
Using these Yukawa coupling constant matrices, we proceed with numerical 
calculations. In evaluating the LFV branching ratios, we first solve 
the RGE's for the soft SUSY breaking parameters and the Yukawa couplings 
in the MSSM to determine the masses and mixings for the SUSY particles. 
Then we input these data into the formulae presented in the previous sections. 

In the following, we list the input parameters what we used for 
the hadronization processes. For the pseudoscalar meson decay constants, 
we take as an input the following values \cite{Eidelman:2004wy},
\bea
f_{\pi^0} &=& 0.119 \,\mathrm{[GeV]}, \;\;
f_{\eta} \ =\ 0.131 \,\mathrm{[GeV]}, \;\;
f_{\eta^\prime} \ =\ 0.118 \,\mathrm{[GeV]}, 
\eea
and for the vector-meson decay constants we use the values, 
extracted from the vector meson decay, $V \to e^+ e^-$,
\bea
\gamma_{\rho^0} &=& 2.518, \;\;
\gamma_{\phi} \ =\ 2.993, \;\;
\gamma_{\omega} \ =\ 3.116. 
\eea
The mixing angles between singlet and octet states 
for the vector mesons and for the pseudoscalar mesons 
that we use are \cite{Eidelman:2004wy},
\bea
\theta_V &=& 35^\circ, \;\;\theta_P \ =\ -17.3^\circ.
\eea
\begin{figure}[!]
   \begin{center}
     \leavevmode
     \epsfxsize=12cm \epsfbox{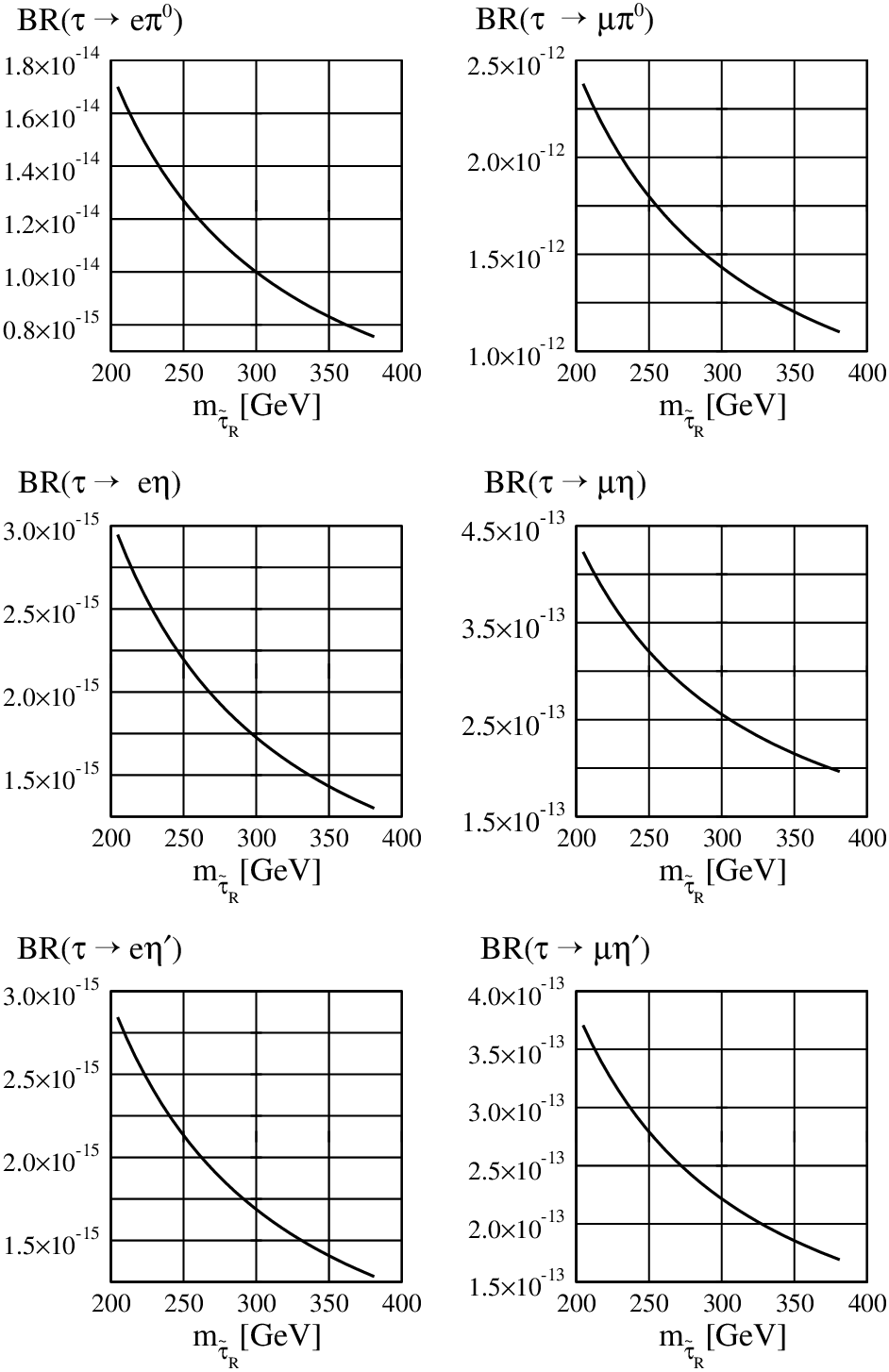}
     \caption{The branching ratios for SL LFV decays 
      $\tau\to e \pi^0/e \eta/e \eta'/\mu \pi^0/\mu \eta/\mu \eta'$
      as a function of mass of lightest charged sfermion $m_{\tilde{\tau}_R}$.
      As an input we have taken $tan\;\beta =45$, $\mu < 0$.
      and $A_0 = 0$. } 
      \label{taulP}
  \end{center}
\end{figure}
\begin{figure}[!]
   \begin{center}
     \leavevmode
     \epsfxsize=12cm \epsfbox{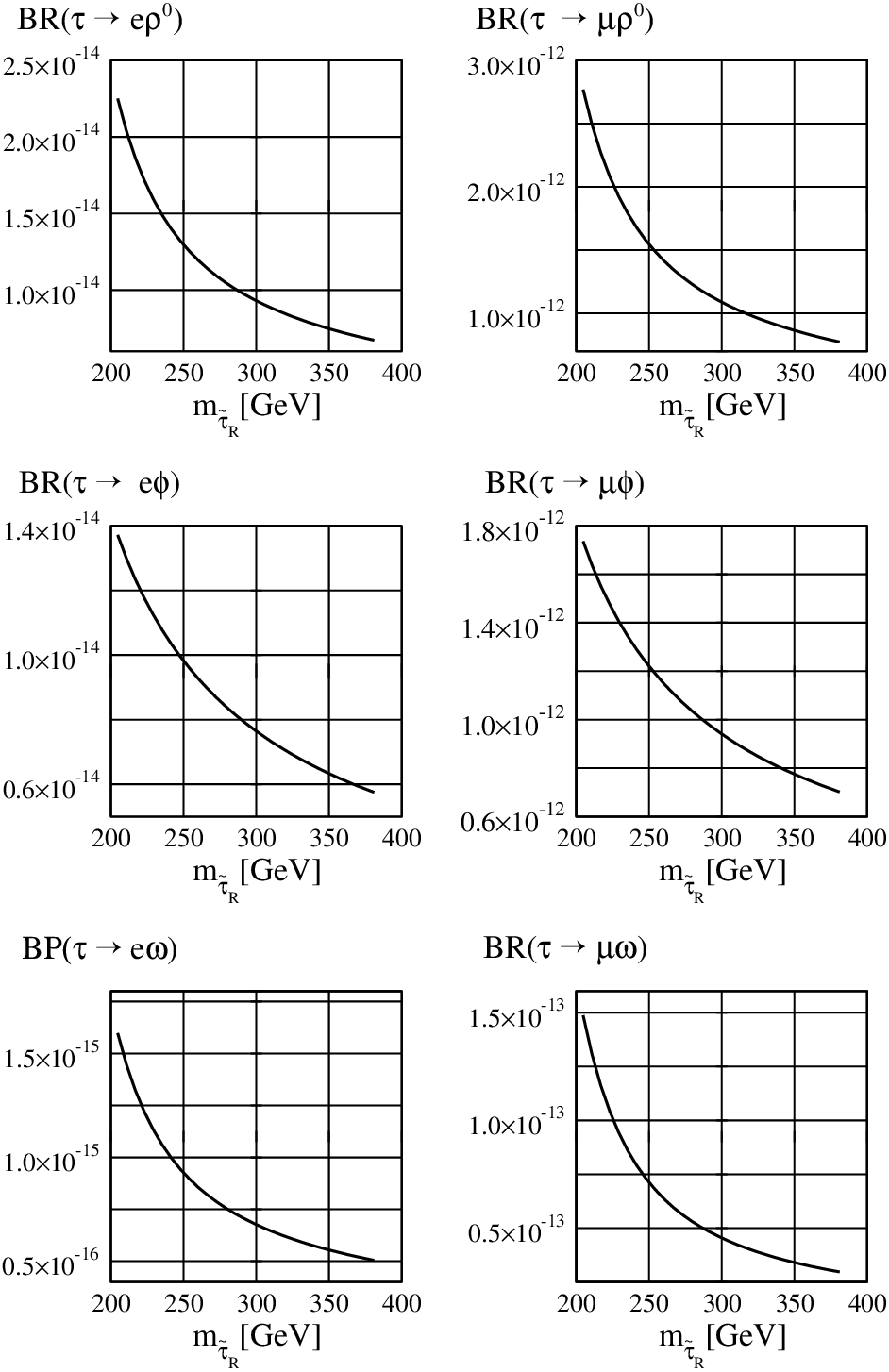}
     \caption{The branching ratios for SL LFV decays
      $\tau\to e \rho^0/e \phi/e \omega/\mu \rho^0/\mu \phi/\mu \omega$
      as a function of mass of lightest charged sfermion $m_{\tilde{\tau}_R}$.
      The input parameters are as in Fig. 1. }
      \label{taulV}
  \end{center}
\end{figure}
\begin{figure}[!]
   \begin{center}
     \leavevmode
     \epsfxsize=12cm \epsfbox{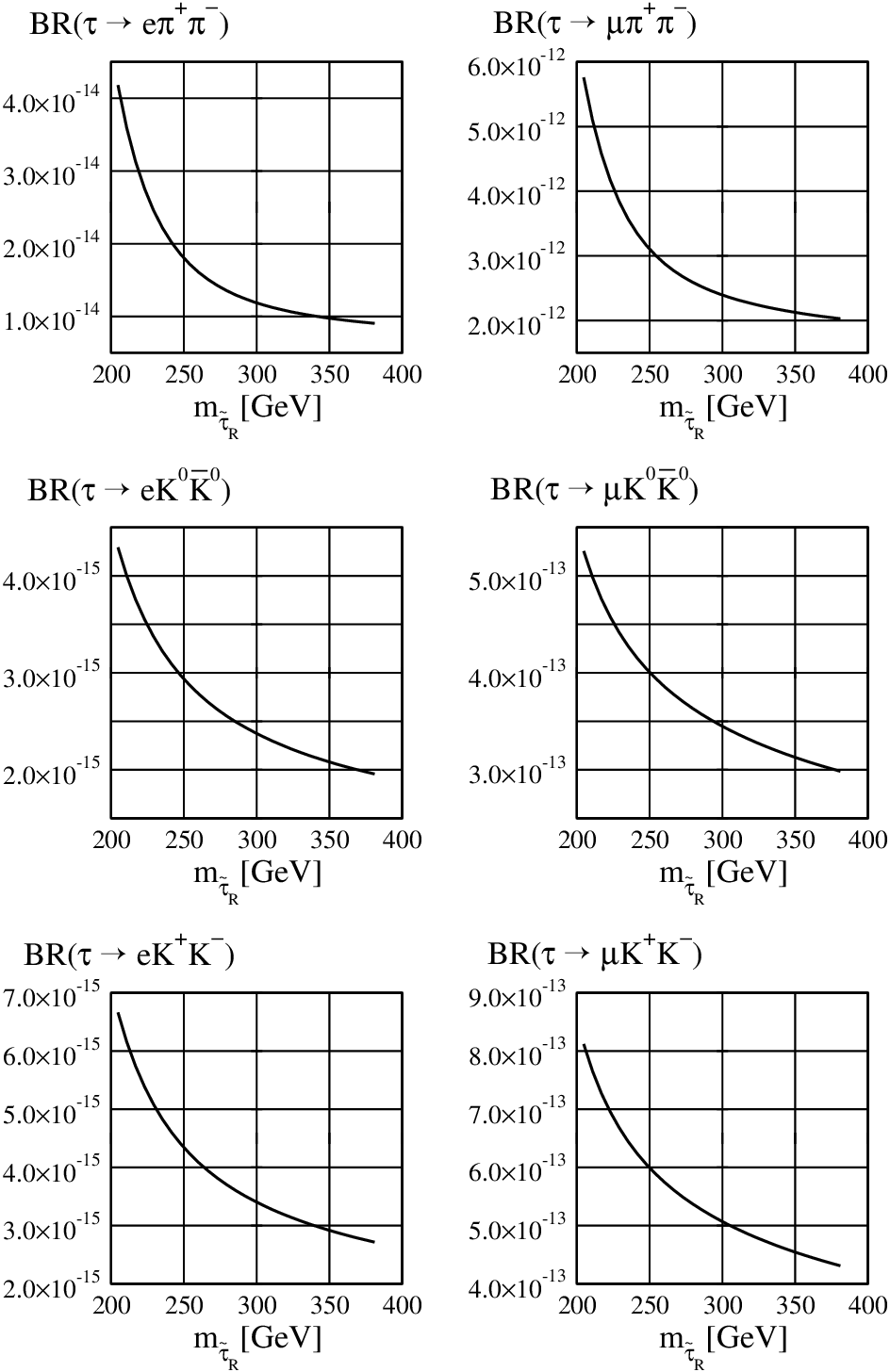}
     \caption{The branching ratios for SL LFV decays
      $\tau\to e \pi^+\pi^-/e K^0\overline{K}^0/e K^+K^-/\mu \pi^+\pi^-/\mu K^+K^-/\mu K^0\overline{K}^0$
      as a function of mass of lightest charged sfermion $m_{\tilde{\tau}_R}$.
      The input parameters are as in Fig. 1. }
      \label{taulPP}
  \end{center}
\end{figure}
\begin{figure}[!]
   \begin{center}
     \leavevmode
     \epsfxsize=12cm \epsfbox{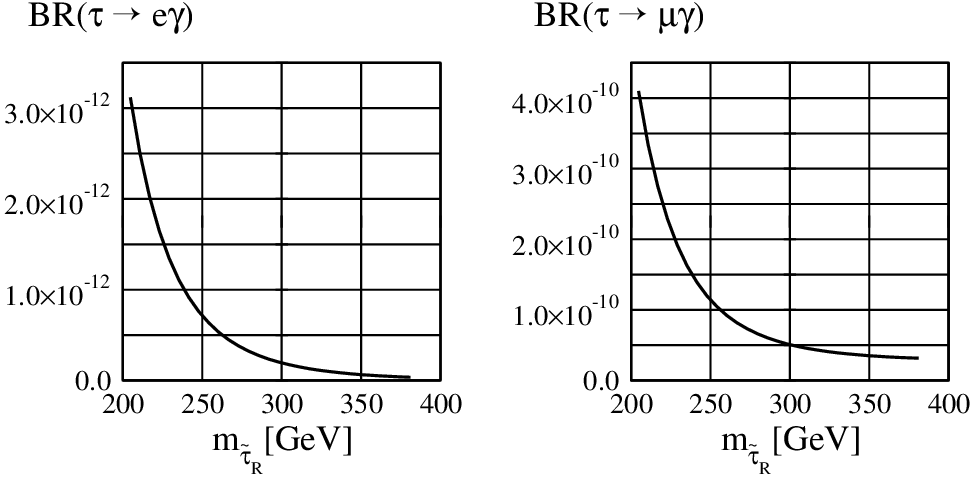}
     \caption{The branching ratios for decays
      $\tau\to e \gamma/\mu \gamma$
      as a function of mass of lightest charged sfermion $m_{\tilde{\tau}_R}$.
      The input parameters are as in Fig. 1. }
      \label{taulg}
  \end{center}
\end{figure}
\begin{figure}[!]
   \begin{center}
     \leavevmode
     \epsfxsize=12cm \epsfbox{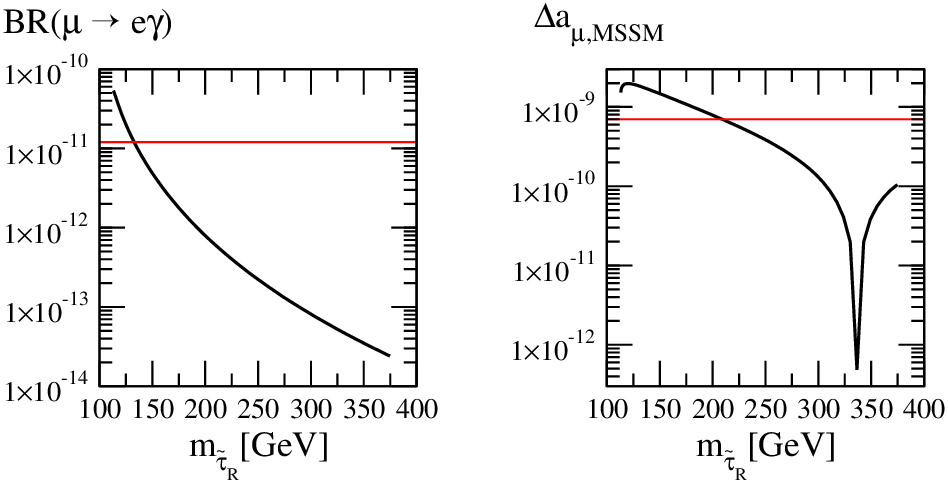}
     \caption{The anomalous magnetic moment and branching ratios for 
      $\mu\to e \gamma$
      as a function of mass of lightest charged sfermion $m_{\tilde{\tau}_R}$.
      These quantities restrict the region of $m_{\tilde{\tau}_R}$ values.
      The present experimental upper limits are represented by horisontal red lines.
      The input parameters are as in Fig. 1. }
      \label{amu_mueg}
  \end{center}
\end{figure}
In order to investigate the dependence of SL LFV branching 
ratios on model parameters we plot their dependence on $m_{\tilde{\tau}_R}$:
$\tau \rightarrow e \pi^0$, $\tau \rightarrow e \eta$, 
$\tau \rightarrow e \eta^\prime$, $\tau \rightarrow \mu \pi^0$, 
$\tau \rightarrow \mu \eta$, and $\tau \rightarrow \mu \eta^\prime$ 
graphs in Fig. 1, 
$\tau \rightarrow e \rho^0$, $\tau \rightarrow \mu \rho^0$
$\tau \rightarrow e \phi$, $\tau \rightarrow \mu \phi$ 
$\tau \rightarrow e \omega$, and $\tau \rightarrow \mu \omega$
graphs in Fig 2,   
$\tau \rightarrow e \pi^+\pi^-$, $\tau \rightarrow \mu \pi^-\pi^+$,
$\tau \rightarrow e K^0\overline{K}^0$, and $\tau \rightarrow \mu K^0\overline{K}^0$,
$\tau \rightarrow e K^+K^-$, and $\tau \rightarrow \mu K^+K^-$,
graphs in Fig 3, and
$\tau \rightarrow e \gamma$ and $\tau \rightarrow \mu \gamma$ 
graphs in Fig 4.
In Fig 5. we plot the $m_{\tilde{\tau}_R}$ dependence of the 
quantities used to determine the alowed region of $m_{\tilde{\tau}_R}$ mass, 
the MSSM contribution to the anomalous magnetic 
moment $\Delta a_\mu$ and  $\mu \rightarrow e \gamma$ branching ratio, 
The parameters are chosen so as to satisfy the WMAP 
constraint on the cold dark matter (CDM) relic density 
\cite{Bennett:2003bz}, 
\be
\Omega_{\mathrm{CDM}} h^2 = 0.1126. 
\ee
We can transmute this value into the approximate relation 
between $m_0$ and $M_{1/2}$, such as 
\be
m_0 \;\mathrm{[GeV]} 
= \frac{9}{28} M_{1/2} \;\mathrm{[GeV]} + 150 \;\mathrm{[GeV]}, 
\ee
for $\tan \beta = 45$, and $A_0=0$. 
The maximal values for the branching ratios 
of the $\tau \rightarrow e/\mu \,\pi^0$ processes 
are found to be 
\bea
\mathrm{BR}(\tau \to e \pi^0) &\simeq& 1.7 \times 10^{-14},
\\
\mathrm{BR}(\tau \to \mu \pi^0) &\simeq& 2.4 \times 10^{-12}, 
\eea
and for the $\tau \rightarrow e/\mu \,\eta$ processes 
$\mathrm{BR}(\tau \to e/\mu \,\eta) \simeq 0.15 \times 
\mathrm{BR}(\tau \to e/\mu \,\pi^0)$, 
with suitably chosen CMSSM parameters which can realize 
the neutralino dark matter scenario by the WMAP data. 
It can be realized with the following set of parameters: 
$\tan \beta=45$, $\mu>0$ and $A_0=0$, with 
$M_{1/2}=600 \;\mathrm{[GeV]}$, $m_0= 343\; \mathrm{[GeV]}$. 
This parameter set can also predict the muon $g-2$ 
within the range of  the recent result of Brookhaven E821 
experiment and also provides the $\tau \to \mu \gamma$ and 
$\mu \rightarrow e \gamma$ branching ratios close to 
the current experimental bound \cite{Fukuyama:2003hn}.  
The ratio between two processes $\tau \rightarrow e/\mu \,\pi^0$ and 
$\tau \rightarrow e/\mu \,\eta$ is a result of the 
dominance of the $Z$-boson-penguin amplitude in these processes, and reflects the difference 
in the form factors and the mixings between the singlet state and 
the octet state of the $\eta$ mesons, 
\be
\frac{\mathrm{BR}(\tau \to e/\mu \, \eta)}
{\mathrm{BR}(\tau \to e/\mu \,\pi)} 
\sim \left(\frac{f_\eta}{f_\pi} \right)^2 
\times \left(\frac{c_P}{\sqrt{3}} + \frac{s_P}{\sqrt{6}} \right)^2 
\sim 0.15. 
\ee
We can also see the correlation between the branching 
ratios for the processes 
$\tau \to e/\mu \,\rho^0$ and $\tau \to e/\mu\, \gamma$ as 
$\mathrm{BR}(\tau \to e/\mu \, \rho^0) \simeq 5.5 \times 10^{-3} \times 
\mathrm{BR}(\tau \to e/\mu \,\gamma)$.  
It can be estimated on the basis 
of the photon-penguin-amplitude dominance in these amplitudes, giving
\be
\frac{\mathrm{BR}(\tau \to e/\mu \, \rho^0)}
{\mathrm{BR}(\tau \to e/\mu \,\gamma)} 
\sim \frac{1}{2} \left(\frac{e}{\gamma_{\rho^0}} \right)^2
\sim 7 \times 10^{-3}. 
\ee
When we impose the constraints from discrepancy of $\tau$ and 
$e^+ e^-$ data in the muon $g-2$ measurements \cite{Bennett:2004pv} 
and from upper limits on the $\mu \to e \gamma$ branching ratio, 
we obtain that the model permits the 
of $m_{\tilde{\tau}_R}$ values satisfying, 
%
%
\be
m_{\tilde{\tau}_R} > 204\;\mathrm{[GeV]} \;. 
\ee
The lower bound comes from the muon $g-2$ constraint. The   
$\mu \to e \gamma$ gives also the lower bound but it is below the
lower bound from the muon $g-2$ constraint. The $g-2$ 
curve has uprising behaviour above $m_{\tilde{\tau}_R} = 340 \;\mathrm{[GeV]}$,
but at $m_{\tilde{\tau}_R} = 1000\;\mathrm{[GeV]}$ it is almost independent
on $m_{\tilde{\tau}_R}$ and has a value $5.4 \times 10^{-10}$, slightly below the present 
experimental $g-2$ uncertainty. Therefore, one can expect that the improvement of the
$g-2$ measurements 
will give the upper limit on $m_{\tilde{\tau}_R}$, too.
Using 
the lower bound  on $m_{\tilde{\tau}_R}$ values one can find 
the theoretical upper bounds for all leptonic and SL LFV 
branching ratios. Leptonic and dominant SL LFV deacays are given in Table I. 
\TABLE[h]{
\begin{tabular}{|c|c|c|}
\hline \hline
Process & Theor. upper bound & Exp. upper bound\\
\hline
$\mu \to e \gamma$          & $6.9 \times 10^{-13}$ & $1.2 \times 10^{-11}$ \\
$\tau \to e \gamma$         & $3.1 \times 10^{-12}$ & $1.0 \times 10^{-9}$ \\
$\tau \to \mu \gamma$       & $4.1 \times 10^{-10}$ & $4.5 \times 10^{-9}$ \\
$\tau \to e \pi^0$         & $1.7 \times 10^{-14}$ & $1.9 \times 10^{-7}$ \\
$\tau \to \mu \pi^0$       & $2.4 \times 10^{-12}$ & $4.3 \times 10^{-7}$ \\
$\tau \to e \eta$          & $2.9 \times 10^{-15}$ & $2.3 \times 10^{-7}$ \\
$\tau \to \mu \eta$        & $4.2 \times 10^{-13}$ & $2.3 \times 10^{-7}$ \\
$\tau \to e \eta^\prime$   & $2.8 \times 10^{-15}$ & $10  \times 10^{-7}$ \\
$\tau \to \mu \eta^\prime$ & $3.7 \times 10^{-13}$ & $4.1 \times 10^{-7}$ \\
$\tau \to e \rho^0$        & $2.3 \times 10^{-14}$ & $2.0 \times 10^{-6}$ \\
$\tau \to \mu \rho^0$      & $2.8 \times 10^{-12}$ & $6.3 \times 10^{-6}$ \\
$\tau \to e \phi$          & $1.4 \times 10^{-14}$ & $6.9 \times 10^{-6}$ \\
$\tau \to \mu \phi$        & $1.7 \times 10^{-12}$ & $7.0 \times 10^{-6}$ \\
$\tau \to e \omega$        & $1.6 \times 10^{-15}$ & $-$ \\
$\tau \to \mu \omega$      & $1.5 \times 10^{-13}$ & $-$ \\
$\tau \to e \pi^+ \pi^-$        & $4.2 \times 10^{-14}$ & $8.7 \times 10^{-7}$\\
$\tau \to \mu\pi^+\pi^-$        & $5.8 \times 10^{-12}$ & $2.8 \times 10^{-7}$\\
$\tau \to e K^0 \overline{K}^0$   & $4.3 \times 10^{-15}$ & $2.2 \times 10^{-6}$ \\
$\tau \to \mu K^0 \overline{K}^0$ & $5.3 \times 10^{-13}$ & $3.4 \times 10^{-6}$ \\
$\tau \to e K^+ K^-$              & $6.7 \times 10^{-15}$ & $3.0 \times 10^{-7}$\\
$\tau \to \mu K^+ K^-$            & $5.1 \times 10^{-13}$ & $11.7 \times 10^{-7}$\\
\hline \hline
\end{tabular}
\caption{Theoretical upper bounds $\ell \to \ell' \gamma$ processes 
and dominant SL LFV processes. The upper bound is obtained from 
the muon $g-2$ constraint. 
We referred the experimental data mainly from \cite{tau04} and partly 
from \cite{Eidelman:2004wy}.}}
\section{Summary} 
The evidence for the neutrino masses and flavour mixings implies 
the non-conservation of the lepton-flavour symmetry. Thus, 
the LFV processes in the charged-lepton sector are expected. 
In supersymmetric model based on the minimal $SO(10)$ model, the values for 
the rates of the LFV processes are generally still several orders 
of magnitudes below the accessible current 
experimental bounds. In this paper, we have presented 
the detailed theoretical description for the SL LFV decays of 
the charged leptons with one or two pseudoscalar mesons or 
one vector meson in the final state. Also, some previous formulae 
have been corrected. The $\gamma$-penguin amplitude is corrected 
to assure the gauge invariance, the Z-penguin amplitude is corrected, 
new box contributions to the box amplitude has been found and 
previously neglected terms are given. 

To evaluate the decay rates of the LFV processes within the MSSM, 
the parameters and the LFV interactions of the MSSM have to be 
specified. It has been shown \cite{Fukuyama:2002ch} 
that the minimal SUSY $SO(10)$ model can simultaneously 
accommodate all the observed quark-lepton mass matrix data involving 
the neutrino oscillation data with appropriately fixed free parameters. 
In this model, the neutrino-Dirac-Yukawa coupling matrix are completely 
determined, and its off-diagonal components are the primary source 
of the lepton-flavour violation in the basis where the charged-lepton 
and the right-handed neutrino mass matrices are real and diagonal. 
Using this Yukawa coupling matrix, we have calculated the rate of 
the LFV processes assuming the mSUGRA scenario. 
The analytical formulae of various SL LFV processes, 
$\ell_i \rightarrow \ell_j P,~\ell_i \rightarrow \ell_j V,~
\ell_i \rightarrow \ell_j P P$ are given. 
Using these formulae, we have numerically evaluated 
$\ell_i \rightarrow \ell_j P,~\ell_i \rightarrow \ell_j V$ and 
$\ell_i \rightarrow \ell_j P_1P_2$ branching ratios. 
Among these, the branching ratios of $\tau\rightarrow \mu \gamma$ and $\mu\rightarrow e
\gamma$ may be interesting for the near future experiments.
The typical CMSSM parameters used in calculations can realize 
the neutralino dark matter scenario by the WMAP data. 

\acknowledgments

A.I. would like to thank to I. Picek and S. Fajfer for 
the discussions on current-quark masses and evaluation of 
tensor-quark current. 
A part of the work was presented by A.I. at the workshop, 
``The 8th International Workshop on Tau-Lepton Physics (Tau04)'', 
held in Nara-ken New Public Hall, Japan. 
We are grateful to all organizers of this workshop and 
particularly to Prof. Ohshima for his kind hospitality extended to 
A.I and T.K. during their stay at Nagoya University. 
T.F. would like to thank S.T. Petcov for his hospitality at SISSA. 
This work of T.F. and T.K. was supported by the Grant in Aid 
for Scientific Research from the Ministry of Education, 
Science and Culture and the work of T.K. was supported by 
the Research Fellowship of the Japan Society for the Promotion 
of Science (\# 16540269 and \# 7336). 
The work of A.I is supported by the Ministry of Science and 
Technology of Republic of Croatia under contract 0119261. 

\appendix
\section{Notation for the MSSM Lagrangian}
Here we summarize our notation necessary for defining the 
masses of the sparticles in the MSSM Lagrangian. 
\be
v \equiv \sqrt{{\langle H^{0}_{u} \rangle}^{2}
                +{\langle H^{0}_{d} \rangle}^{2}} = 174.1 \;\mathrm{[GeV]}, 
\ee
and
\be
\tan\beta \equiv \frac{\langle H^{0}_{u} \rangle}{\langle H^{0}_{d} \rangle}.
\ee
Then the charged fermion mass matrices are given by
\bea
M_{u}^{ij} &=& - Y_{u}^{ij} v \sin \beta, 
\\
M_{d}^{ij} &=& Y_{d}^{ij} v \cos \beta, 
\\
M_{e}^{ij} &=& Y_{e}^{ij} v \cos \beta. 
\eea
The mass matrix of the charginos is written as: 

\begin{eqnarray}
&& {\cal L} = 
-\left(
\begin{array}{ll}
\overline{\widetilde{W}^{-}_{R}}, &
\overline{\widetilde{H}^{-}_{u\,R}}  
\end{array} 
\right) \,
{M}_{\tilde{\chi}^{\pm}}\,
\left(
\begin{array}{l}
\widetilde{W}^{-}_{L} \\
\widetilde{H}_{d\,L}^{-}
\end{array}
\right) + h.c. ,
 \nonumber \\
&& {M}_{\tilde{\chi}^{\pm}} = 
\left(
\begin{array}{cc}
M_{2} & \sqrt{2}M_{W}\cos\beta \\
\sqrt{2}M_{W}\sin\beta & \mu 
\end{array}
\right). 
\end{eqnarray}
The mass matrix of the neutralinos is written as: 
\begin{eqnarray}
&& {\cal L} = -\frac{1}{2}
\left(
\begin{array}{llll}
{\widetilde{B}_L}, & 
{\widetilde{W}^{3}_{L}}, &
{\widetilde{H}^{0}_{d\,L}}, & 
{\widetilde{H}^{0}_{u\,L}}
\end{array} 
\right)\,
{M}_{\tilde{\chi}^{0}} \,
\left(
\begin{array}{l}
\widetilde{B}_{L} \\
\widetilde{W}^{3}_{L}\\
\widetilde{H}_{d\,L}^{0} \\ 
\widetilde{H}_{u\,L}^{0}
\end{array}
\right) + h.c. ,
 \nonumber \\
&& {M}_{\tilde{\chi}^{0}} = 
 \left(
\begin{array}{cccc}
M_{1} & 0     & -M_{Z}\sin\theta_{W}\cos\beta
      & M_{Z}\sin\theta_{W}\sin\beta \\
0     & M_{2} & M_{Z}\cos\theta_{W}\cos\beta
      & -M_{Z}\cos\theta_{W}\sin\beta \\
        -M_{Z}\sin\theta_{W}\cos\beta & M_{Z}\cos\theta_{W}\cos\beta
      & 0 & -\mu \\
        M_{Z}\sin\theta_{W}\sin\beta & -M_{Z}\cos\theta_{W}\sin\beta
      & -\mu & 0
\end{array}
\right). 
\nonumber \\
\end{eqnarray}
The mass matrices of the squarks are written as follows:
\begin{eqnarray}
{\cal L} &=&-(m_{\tilde{u}}^{2})^{ij}\;
\tilde{u}^{\dagger}_i
\tilde{u}_j
-(m_{\tilde{d}}^{2})^{ij}\;
\tilde{d}^{\dagger}_i
\tilde{d}_j, \nonumber \\
m_{\tilde{u}}^{2} &=& 
\left(
\begin{array}{cc}
m^{2}_{\tilde{q}} + M_{u}^{\dag}M_{u} & 
-A_u^{\dag} v \sin\beta  - M_u^{\dagger}\; \mu^{*} \cot\beta \\
-A_u v \sin\beta - M_u \;\mu \cot\beta & 
m^{2}_{\tilde{u}} + M_{u}M_{u}^{\dag}
-M_{Z}^{2}\cos 2\beta\sin^{2}\theta_{W}
\end{array}
\right)
\nonumber\\
&+& 
\left(
\begin{array}{cc}
M_{Z}^{2}\cos 2\beta(\frac{1}{2}-\frac{2}{3}\sin^{2}\theta_{W})
{\bf 1}_{3\times3}
& 0 \\
0 & 
\frac{2}{3}M_{Z}^{2}\cos 2\beta \sin^{2}\theta_{W} {\bf 1}_{3\times3}
\end{array}
\right), 
\nonumber\\
m_{\tilde{d}}^{2} &=&
\left(
\begin{array}{cc}
m^{2}_{\tilde{q}} + M_{d}^{\dag}M_{d} & 
A_d^{\dag} v \cos\beta  - M_d^{\dagger} \;\mu^{*} \tan\beta \\
A_d v \cos\beta - M_d \;\mu \tan\beta & 
m^{2}_{\tilde{d}} + M_{d}M_{d}^{\dag}
-M_{Z}^{2}\cos 2\beta\sin^{2}\theta_{W}
\end{array}
\right)
\nonumber\\
&+& 
\left(
\begin{array}{cc}
M_{Z}^{2}\cos 2\beta(-\frac{1}{2}+\frac{1}{3}\sin^{2}\theta_{W})
{\bf 1}_{3\times3}
& 0 \\
0 & 
-\frac{1}{3} M_{Z}^{2}\cos 2\beta \sin^{2}\theta_{W} {\bf 1}_{3\times3}
\end{array}
\right).
\end{eqnarray}
The mass matrices of the sleptons are written as follows:
\begin{eqnarray}
{\cal L} &=&
-(m_{\tilde{\nu}}^{2})^{ij}\;
\tilde{\nu}^{\dagger}_i
\tilde{\nu}_j
-(m_{\tilde{e}}^{2})^{ij}\;
\tilde{e}^{\dagger}_i
\tilde{e}_j,
 \nonumber \\
m_{\tilde{\nu}}^{2} &=& 
m^{2}_{\tilde{\ell}} +\frac{1}{2}M_{Z}^{2}\cos 2\beta 
{\bf 1}_{3 \times 3}, 
\nonumber\\
m_{\tilde{e}}^{2} &=&
\left(
\begin{array}{cc}
m^{2}_{\tilde{\ell}} + M_{e}^{\dag}M_{e} & 
A_e^{\dag} v \cos\beta  - M_e^{\dagger}\; \mu^{*} \tan\beta \\
A_e v \cos\beta - M_e \;\mu \tan\beta & 
m^{2}_{\tilde{e}} + M_{e}M_{e}^{\dag}
-M_{Z}^{2}\cos 2\beta\sin^{2}\theta_{W}
\end{array}
\right)
\nonumber\\
&+& 
\left(
\begin{array}{cc}
M_{Z}^{2}\cos 2\beta(-\frac{1}{2}+\sin^{2}\theta_{W})
{\bf 1}_{3\times3}
& 0 \\
0 & 
-M_{Z}^{2}\cos 2\beta \sin^{2}\theta_{W} {\bf 1}_{3\times3}
\end{array}
\right).
\end{eqnarray}
They are diagonalized with unitary matrices as follows: 
\bea
O_R \, M_{\tilde{\chi}^{\pm}} \, O_L^{\dagger} 
&=&
{\mathrm{diag}} 
(M_{\widetilde{\chi}_{1}^{-}}, M_{\widetilde{\chi}_{2}^{-}}),
\nonumber\\
O_N^* \, M_{\tilde{\chi}^{0}} \, O_N^{\dagger} 
&=&
{\mathrm{diag}} 
(M_{\widetilde{\chi}_{1}^{0}}, M_{\widetilde{\chi}_{2}^{0}}, 
M_{\widetilde{\chi}_{3}^{0}}, M_{\widetilde{\chi}_{4}^{0}}),
\nonumber\\
U_{\widetilde f} \, m_{\widetilde{f}}^2 \, U_{\widetilde f}^\dagger 
&=&
{\mathrm{diag}} 
(m_{\widetilde{f}_{1}}^2,....,m_{\widetilde{f}_{6}}^2), 
\; (f \ =\ u, d, e), 
\nonumber\\
U_{\widetilde \nu} \, m_{\widetilde{\nu}}^2 \, U_{\widetilde \nu}^\dagger 
&=&
{\mathrm{diag}} 
(m_{\widetilde{\nu}_{1}}^2, m_{\widetilde{\nu}_{2}}^2,
m_{\widetilde{\nu}_{3}}^2). 
\eea

\section{Lagrangian for fermion-sfermion-gaugino/Higgsino 
interactions in MSSM}
The LFV interactions in the MSSM include 
the fermion-sfermion-gaugino/Higgsino vertices. 
These vertices, and corresponding coupling constants 
($C_{iAX}^{L,R(f)}$, $N_{iAX}^{L,R(f)}$, $f=\nu,e,u,d$) 
are defined by the following Lagrangian 
\bea
{\cal L}
&=&
\bar{u_i}
\left[C_{iAX}^{L(u)}P_{L}+C_{iAX}^{R(u)}P_{R} \right]
\widetilde{\chi}_A^{+} \widetilde{d}_{X}
+
\bar{d_i}
\left[C_{iAX}^{L(d)}P_{L}+C_{iAX}^{R(d)}P_{R} \right]
\widetilde{\chi}_A^{-} \widetilde{u}_{X}
\nonumber\\
&+&
\bar{\nu_i} C_{iAX}^{R(\nu)}P_{R}
\widetilde{\chi}_A^{+} \widetilde{e}_{X}
+
\bar{e_i}
\left[C_{iAX}^{L(e)}P_{L}+C_{iAX}^{R(e)}P_{R} \right]
\widetilde{\chi}_A^{-} \widetilde{\nu}_{X}
\nonumber\\
&+&
\bar{u_i}
\left[N_{iAX}^{L(u)}P_{L}+N_{iAX}^{R(u)}P_{R} \right]
\widetilde{\chi}_A^{0} \widetilde{u}_{X}
+
\bar{d_i}
\left[N_{iAX}^{L(d)}P_{L}+N_{iAX}^{R(d)}P_{R} \right]
\widetilde{\chi}_A^{0} \widetilde{d}_{X}
\nonumber\\
&+&
\bar{\nu_i} N_{iAX}^{R(\nu)}P_{R}
\widetilde{\chi}_A^{0} \widetilde{\nu}_{X}
+
\bar{e_i}
\left[N_{iAX}^{L(e)}P_{L}+N_{iAX}^{R(e)}P_{R} \right]
\widetilde{\chi}_A^{0} \widetilde{e}_{X}
\nonumber\\
&+&h.c.
\nonumber\\
&\equiv&
\overline{u}_{i} P_L \tilde{\chi}^+_{A} \tilde{d}_X
\left[g \left\{ - \frac{m_{u_i}}{\sqrt{2} M_{\mathrm{W}} \sin \beta}
(O_{R})_{A 2} (U_{\tilde d}^*)_{Xi}  \right\} \right]
\nonumber\\
&+&
\overline{u}_{i} P_R \tilde{\chi}^+_{A} \tilde{d}_X
\left[g \left\{(O_{L})_{A1} (U_{\tilde d}^*)_{Xi} 
\ +\ \frac{m_{d_k}}{\sqrt{2} M_{\mathrm{W}} \cos \beta} 
(V_{\mathrm{CKM}})_{kj}(V_{\mathrm{CKM}}^*)_{ki}
(O_{L})_{A 2} (U_{\tilde d}^*)_{X,j+3} \right\} \right]
\nonumber\\
&+&
\overline{d}_{i} P_L \tilde{\chi}^-_{A} \tilde{u}_X
\left[ g \left\{\frac{m_{d_i}}{\sqrt{2} M_{\mathrm{W}} \cos \beta}
(O_{L}^*)_{A 2} (U_{\tilde u}^*)_{Xj} \right\} 
(V_{\mathrm{CKM}})_{ij} \right]
\nonumber\\
&+&
\overline{d}_{i} P_R \tilde{\chi}^-_{A} \tilde{u}_X
\left[ g \left\{(O_{R}^*)_{A1} (U_{\tilde u}^*)_{Xj}
\ -\ \frac{m_{u_j}}{\sqrt{2} M_{\mathrm{W}} \sin \beta}
(O_{R}^*)_{A 2} (U_{\tilde u}^*)_{X,j+3} \right\}
(V_{\mathrm{CKM}}^{*})_{ij} \right]
\nonumber\\
&+&
\overline{\nu}_{i} P_R \tilde{\chi}^+_{A} \tilde{e}_X
\left[ g \left\{(O_{L})_{A1} (U_{\tilde e}^*)_{Xj} 
\ +\ \frac{m_{e_i}}{\sqrt{2} M_{\mathrm{W}} \cos \beta}
(O_{L})_{A 2} (U_{\tilde e}^*)_{X,j+3} \right\}
(U_{\mathrm{MNS}}^{*})_{ij} \right]
\nonumber\\
&+&
\overline{e}_{i} P_L \tilde{\chi}^-_{A} \tilde{\nu}_X
\left[ g \left\{ \frac{m_{e_i}}{\sqrt{2} M_{\mathrm{W}} \cos \beta}
(O_{L}^*)_{A 2} (U_{\tilde \nu}^*)_{Xi} \right\} \right]
\nonumber\\
&+&
\overline{e}_{i} P_R \tilde{\chi}^-_{A} \tilde{\nu}_X
\Bigg[ g \left\{(O_{R}^*)_{A1} (U_{\tilde \nu}^*)_{Xi} \right\} \Bigg]
\nonumber\\
&+&
\overline{u}_{i} P_L \tilde{\chi}^0_A \tilde{u}_X
\left[ \frac{g}{\sqrt{2}} \left\{
\frac{m_{u_i}}{M_{\mathrm{W}} \sin \beta}
(O_{N}^*)_{A4} (U_{\tilde u}^*)_{Xi} 
\ -\ \frac{4}{3} \tan \theta_{W} (O_{N}^{*})_{A1}
(U_{\tilde u}^{*})_{X,i+3} \right\} \right]
\nonumber\\
&+&
\overline{u}_{i} P_R \tilde{\chi}^0_A \tilde{u}_X
\left[ \frac{g}{\sqrt{2}} \left\{
\frac{m_{u_i}}{M_{\mathrm{W}} \sin \beta}
(O_{N})_{A4} (U_{\tilde u}^*)_{X,i+3} 
\ +\ \left[(O_{N})_{A2} +
\frac{1}{3}\tan \theta_{W} (O_{N})_{A1} \right]
(U_{\tilde u}^*)_{Xi}  \right\} \right]
\nonumber\\
&+&
\overline{d}_{i} P_L \tilde{\chi}^0_A \tilde{d}_X
\left[ \frac{g}{\sqrt{2}} \left\{
- \frac{m_{d_i}}{M_{\mathrm{W}} \cos \beta}
(O_{N}^*)_{A3} (U_{\tilde d}^*)_{Xj}
\ +\ \frac{2}{3} \tan \theta_{W} (O_{N}^{*})_{A1}
(U_{\tilde d}^{*})_{X,j+3}  \right\} 
(V_{\mathrm{CKM}})_{ij} \right]
\nonumber\\
&+&
\overline{d}_{i} P_R \tilde{\chi}^0_A \tilde{d}_X
\left[ \frac{g}{\sqrt{2}} \left\{
- \frac{ m_{d_i}}{M_{\mathrm{W}} \cos \beta}
(O_{N})_{A3} (U_{\tilde d}^*)_{X,j+3} 
\right. \right.
\nonumber\\
&& \left. \left.+ \left[-(O_{N})_{A2}+
\frac{1}{3} \tan \theta_{W} (O_{N})_{A1} \right]
(U_{\tilde d}^*)_{Xj}  \right\}
(V_{\mathrm{CKM}})_{ij} \right]
\nonumber\\
&+&
\overline \nu_{i} P_R \tilde{\chi}^0_A \tilde{\nu}_X
\left[ \frac{g}{\sqrt{2}}
\left[(O_{N})_{A2} -
\tan \theta_{W} (O_{N})_{A1}  \right]
(U_{\tilde \nu}^*)_{X,j}
(U_{\mathrm{MNS}})_{ij} \right]
\nonumber\\
&+&
\overline{e}_{i} P_L \tilde{\chi}^0_A \tilde{e}_X
\left[ \frac{g}{\sqrt{2}} \left\{
-\frac{m_{e_i}}{M_{\mathrm{W}} \cos \beta}
(O_{N}^*)_{A3} (U_{\tilde e}^*)_{Xi}
\ +\ 2 \tan \theta_{W} (O_{N}^{*})_{A1}
(U_{\tilde e}^{*})_{X,i+3} \right\} \right]
\nonumber\\
&+&
\overline{e}_{i} P_R \tilde{\chi}^0_A \tilde{e}_X
\left[ \frac{g}{\sqrt{2}} \left\{
- \frac{m_{e_i}}{M_{\mathrm{W}} \cos \beta}
(O_{N})_{A3} (U_{\tilde e}^*)_{X,i+3} 
\ +\ \left[-(O_{N})_{A2}
- \tan \theta_{W} (O_{N})_{A1} \right]
(U_{\tilde e}^*)_{Xi} \right\} \right]
\nonumber\\
&+& h.c.
\eea

\section{Trilinear interactions of fermions or bosons 
with $Z^0$-boson or photon} 
The interactions of $Z^0$-boson or photon with any fermion 
or any boson follow from $SU(2)_L\times U(1)_Y$ gauge symmetry. 
\bea
{\cal L}_f &=& -g \sum_f
\overline{f}
\left[
\Slash{Z} 
\left( I^f_3\frac{1}{c_W} - Q^f \frac{s_W^2}{c_W} \right)
+ 
\Slash{A}
\left( s_W Q^f \right)
\right]f,
\\
{\cal L}_b &=& -g \sum_b
\left( b^\dagger\:
i\mbox{\hspace{-.3em}} \stackrel{\leftrightarrow}{\partial}_\mu
b \right)
\left[
Z^\mu \left( I^f_3\frac{1}{c_W} - Q^f \frac{s_W^2}{c_W} \right)
+ A^\mu \left( s_W Q^f \right)
\right]. 
\eea
The interaction Lagrangians are flavour-diagonal in the weak basis. 
In the mass basis the interactions with photon remain diagonal, 
because all mixed states $A$ must have the same charge, $Q_A$. 
On the other side the interactions with $Z$-boson, which depend on 
charge ($Q_A$) and third component of the weak isospin ($I_{3A}$) 
are not in general flavour-diagonal in the mass basis, 
because the mixed states may have different $I_{3A}$ values.

For LFV processes the interaction of photon and $Z$-boson with 
charginos, neutralinos and sfermion fields is needed. 
The Lagrangians for the corresponding weak-basis fields 
is easily written knowing the charges and $I_3$-s of these fields. 
After transformation from the weak-basis to the mass-basis 
the following interaction Lagrangians are obtained 
\bea
{\cal L}_{\chi^-} &=&
-g \overline{\tilde{\chi}^-_A}
\Slash{Z}
\Big[ 
(E_{AB}^{L(\chi^-)} P_L + E_{AB}^{R(\chi^-)} P_R )
+\frac{s_W^2}{c_W} \delta_{BA} \Big]
\tilde{\chi}^-_B
+
e \overline{\tilde{\chi}^-_A} \Slash{A} 
\delta_{AB} \tilde{\chi}^-_B,
\\
{\cal L}_{\chi^0} &=&
-g \overline{\tilde{\chi}^0_A}
\Slash{Z}
\Big[
E_{AB}^{L(\chi^0)} P_L + E_{AB}^{R(\chi^0)} P_L 
\Big]
\tilde{\chi}^0_B,
\\
{\cal L}_{\tilde{f}} &=&
-g 
\tilde{f}^*_X
i\mbox{\hspace{-.3em}} 
\stackrel{\leftrightarrow}{\partial}_\mu \mbox{\hspace{-.3em}}
\tilde{f}_Y 
Z^\mu\
[D_{XY}^{\tilde{f}}]
-
e \tilde{f}^*_X
i\mbox{\hspace{-.3em}} 
\stackrel{\leftrightarrow}{\partial}_\mu \mbox{\hspace{-.3em}}
\tilde{f}_Y
A^\mu\
[Q^{\tilde{f}}],
\eea
where
\bea
E_{AB}^{L(\chi^-)} &=& 
-\frac{1}{c_W} \bigg( (O_L)_{A1} (O_L^*)_{B1} 
+ \frac{1}{2} (O_L)_{A2} (O_L^*)_{B2} \bigg), 
\\
E_{AB}^{R(\chi^-)} &=&
-\frac{1}{c_W} \bigg( (O_R)_{A1} (O_R^*)_{B1} 
+ \frac{1}{2} (O_R)_{A2} (O_R^*)_{B2} \bigg), 
\\
E_{AB}^{L(\chi^0)} &=&
-\frac{1}{c_W} \bigg( \frac{1}{2} (O_N)_{A3} (O_N^*)_{B3}  
  - \frac{1}{2} (O_N)_{A4} (O_N^*)_{B4} \bigg),
\\
E_{AB}^{R(\chi^0)} &=& - E_{AB}^{L(\chi^0)*},
\\
D^{\tilde{f_L}}_{XY} &=&
I^{\tilde{f_L}}_3 (U_{\tilde{f}})_{Xi} 
(U^*_{\tilde{f}})_{Yi} - Q^{\tilde{f_L}}_3 \frac{s_W^2}{c_W}\;.
\eea
Here, in definition of $D_{XY}^{\tilde{f}}$ 
index $i$ is summed over generation indices $1,2,3$.  
$Q^{\tilde{f_L}}$ is a charge of the sfermion $\tilde{f_L}$, 
and $I^{\tilde{f_L}}_3$ is its third component of isospin 
in the weak basis. The charges and third components of 
the isospin of the weak-basis fields for charginos and 
neutralinos are explicitly written in the definitions of
constants $E_{AB}^{L,R(\chi^-)}$ and $E_{AB}^{L,R(\chi^0)}$.

\section{Loop functions}
In this Appendix the loop functions appearing in the $Z$-boson 
amplitude and the box amplitudes are listed. 

\section*{Z-boson loop functions}
The $Z$-boson amplitude comprises two-loop functions from the 
triangle-diagram part of the amplitude, $F_1(a,b,c)$ and $F_2(a,b,c)$, 
and two-loop functions from the self-energy part of the amplitude,
$f_1(a,b)$ and $f_2(a,b)$ are
\bea
F_1(a,b,c) &=&
-\frac{1}{b-c}
\Bigg[
\frac{a \ln a - b \ln b}{a-b}
-
\frac{a \ln a - c \ln c}{a-c}
\Bigg],
\\
F_2(a,b,c) &=&
\frac{3}{8}
-
\frac{1}{4} \frac{1}{b-c}
\Bigg[
\frac{a^2 \ln a - b^2 \ln b}{a-b}
-
\frac{a^2 \ln a - c^2 \ln c}{a-c}
\Bigg],
\\
f_1(a,b) &=& \frac{1}{2} - \frac{\ln a}{2}
+ \frac{ -a^2 + b^2 + 2 b^2 (\ln a - \ln b) }{ 4 (a-b)^2 },
\\
f_2(a,b) &=& \frac{1}{2} - \frac{\ln b}{2}
+ \frac{ a^2 - b^2 + 2 a^2 (\ln b - \ln a) }{ 4 (a-b)^2 }.
\eea
These loop functions are evaluated by neglecting the momenta of 
incoming and outgoing particles. The functions $F_1$ and $F_2$ 
are symmetric regarding replacement of their arguments ($a$, $b$, $c$). 
In the limit of two equal argument the functions $F_1$ and $F_2$ 
have the following form, 
\bea
F_1(a,b,b) &=& \frac{a - b - a \ln a + a \ln b}{ (a-b)^2 },
\\
F_2(a,b,b) &=& \frac{1}{4} - \frac{ \ln b }{4}
+ \frac{ a^2 - b^2 + 2 a^2 (\ln b - \ln a) }{ 8 (a-b)^2}.
\eea
The arguments of logarithms appearing in the $F_1$ and $F_2$ 
can be divided by any constant, what can be used to redefine 
these functions as functions of two variables, for instance 
$b/a$ and $c/a$. The unpleasant $\ln b$ term in $f_2$ function 
can be replaced with $\ln (b/a)$ because of unitarity cancellations 
in the sum $N^{R(e)}_{jBY}  N^{R(e)*}_{iAX}$, and therefore $f_1$ 
and $f_2$ can be expressed in terms of one variable ($b/a$) only.\\

\section*{Box-loop functions}
Box amplitude contains two-loop functions, $d_0$ and $d_2$. 
\bea
d_0 (x,y,z,w)
&=&
\frac{x \ln x}{(y-x)(z-x)(w-x)}
+
\frac{y \ln y}{(x-y)(z-y)(w-y)}
\nonumber\\
&+&
\frac{z \ln z}{(x-z)(y-z)(w-z)}
+
\frac{w \ln w}{(x-w)(y-w)(z-w)},
\\
d_2 (x,y,z,w)
&=&
\frac{1}{4}\left\{
\frac{x^2 \ln x}{(y-x)(z-x)(w-x)}
+
\frac{y^2 \ln y}{(x-y)(z-y)(w-y)}
\right.
\nonumber\\
&+&\left.
\frac{z^2 \ln z}{(x-z)(y-z)(w-z)}
+
\frac{w^2 \ln w}{(x-w)(y-w)(z-w)}
\right\}.
\eea
As mentioned before, they are also by evaluated neglecting 
the momenta of incoming and outgoing particles.

\section{Meson states and quark currents}
Meson states are assumed to contain valence quarks only. 
The quark-antiquark ($q_aq_b^c$) content of the pseudoscalar-meson 
states is given in the table below. The quark-antiquark content of 
the vector-meson states is obtained replacing the fields 
$K^+$, $K^0$, $\pi^+$, $\pi^0$, $\pi^-$, $\bar{K}^0$ $K^-$, 
$\eta_8$, $\eta_1$, $\eta$ and $\eta'$ by fields 
$K^{*+}$, $K^{*0}$, $\rho^+$, $\rho^0$, $\rho^-$, 
$\overline{K^{*0}}$ $\overline{K^{*-}}$,
$\phi_8$, $\phi_1$, $\phi$ and $\omega$, and the angle $\theta_P$ 
by the angle $\theta_V$. 
\TABLE[h]{
\begin{tabular}{l|ll}\hline
$|M\rangle$ & quark content of $|M\rangle$ & quark content of $M(x)$\\
\hline\hline
$|K^+\rangle$ & $us^c\sim b^\dagger_ud^\dagger_s$ & $su^c\sim d_sb_u$\\
$|K^0\rangle$ & $ds^c$ & $sd^c$\\
$|\pi^+\rangle$ & $ud^c$ & $du^c$\\
$|\pi^0\rangle$ & $\frac{1}{\sqrt{2}}(uu^c-dd^c)$ &
    $\frac{1}{\sqrt{2}}(uu^c-dd^c)$\\
$|\pi^-\rangle$ & $du^c$ & $ud^c$\\
$|K^-\rangle$ & $su^c$ & $us^c$\\
$|\bar{K}^0\rangle$ & $sd^c$ & $ds^c$\\
$|\eta_8\rangle$ & $\frac{1}{\sqrt{6}}(uu^c+dd^c-2ss^c)$ &
    $\frac{1}{\sqrt{6}}(uu^c+dd^c-2ss^c)$\\
$|\eta_1\rangle$ & $\frac{1}{\sqrt{6}}(uu^c+dd^c+ss^c)$ &
    $\frac{1}{\sqrt{6}}(uu^c+dd^c+ss^c)$\\ \hline
$|\eta\rangle$ & $c_P|\eta_8\rangle
-s_P|\eta_1\rangle$ &
    $c_P\eta_8(x)-s_P\eta_1(x)$\\
$|\eta'\rangle$ & $s_P|\eta_8\rangle
+c_P|\eta_1\rangle$ &
    $s_P\eta_8(x)+c_P\eta_1(x)$\\ \hline
\end{tabular}
\caption{Quark content of the pseudoscalar-meson states 
and fields: The meson states listed in the Table~2 correspond 
to the tensor description of meson states \cite{Ilakovac:1995wc}. 
The shorthand notation, $c_P=\cos\theta_P$ and $s_P=\sin\theta_P$ is used.}}
From the quark content of the meson fields one can find the meson 
content of e.g. axial vector (A) and vector (V) quark currents, 
(factor of proportionality, Lorentz indices and 
spinor structures are neglected), 
\bea
(\bar{u}\: u)_A &\sim&
\left(\frac{c_P}{\sqrt{6}}-\frac{s_P}{\sqrt{3}}\right) \eta^\dagger +
\left(\frac{s_P}{\sqrt{6}}+\frac{c_P}{\sqrt{3}}\right) \eta^{'\dagger} +
\frac{1}{\sqrt{2}} \pi^{0\dagger}\ \equiv\
k_{\bar{u}u}^\eta \eta^\dagger +
k_{\bar{u}u}^{\eta'} \eta'^{\dagger} +
k_{\bar{u}u}^{\pi^{0}} \pi^{0\dagger}
\nonumber\\
(\bar{d}\: d)_A &\sim&
\left(\frac{c_P}{\sqrt{6}}-\frac{s_P}{\sqrt{3}}\right) \eta^\dagger +
\left(\frac{s_P}{\sqrt{6}}+\frac{c_P}{\sqrt{3}}\right) \eta'^\dagger -
\frac{1}{\sqrt{2}} \rho^{0\dagger}\ \equiv\
k_{\bar{d}d}^\eta \eta^\dagger +
k_{\bar{d}d}^{\eta'} \eta'^\dagger +
k_{\bar{d}d}^{\pi^{0}} \pi^{0\dagger}
\nonumber\\
(\bar{s}\: s)_A &\sim&
\left(-\frac{2c_P}{\sqrt{6}}-\frac{s_P}{\sqrt{3}}\right) \eta^\dagger +
\left(-\frac{2s_P}{\sqrt{6}}+\frac{c_P}{\sqrt{3}}\right) \eta'^\dagger
\ \equiv\
k_{\bar{s}s}^\eta \eta^\dagger +
k_{\bar{s}s}^{\eta'} \eta'^\dagger
\\
(\bar{u}\: u)_V &\sim&
\left(\frac{c_V}{\sqrt{6}}-\frac{s_V}{\sqrt{3}}\right) \phi^\dagger +
\left(\frac{s_V}{\sqrt{6}}+\frac{c_V}{\sqrt{3}}\right) \omega^\dagger +
\frac{1}{\sqrt{2}} \rho^{0\dagger}\ \equiv\
k_{\bar{u}u}^\phi \phi^\dagger +
k_{\bar{u}u}^\omega \omega^\dagger +
k_{\bar{u}u}^{\rho^{0}} \rho^{0\dagger}
\nonumber\\
(\bar{d}\: d)_V &\sim&
\left(\frac{c_V}{\sqrt{6}}-\frac{s_V}{\sqrt{3}}\right) \phi^\dagger +
\left(\frac{s_V}{\sqrt{6}}+\frac{c_V}{\sqrt{3}}\right) \omega^\dagger -
\frac{1}{\sqrt{2}} \rho^{0\dagger}\ \equiv\
k_{\bar{d}d}^\phi \phi^\dagger +
k_{\bar{d}d}^\omega \omega^\dagger +
k_{\bar{d}d}^{\rho^{0}} \rho^{0\dagger}
\nonumber\\
(\bar{s}\: s)_V &\sim&
\left(-\frac{2c_V}{\sqrt{6}}-\frac{s_V}{\sqrt{3}}\right) \phi^\dagger +
\left(-\frac{2s_V}{\sqrt{6}}+\frac{c_V}{\sqrt{3}}\right) \omega^\dagger
\ \equiv\
k_{\bar{s}s}^\phi \phi^\dagger +
k_{\bar{s}s}^\omega \omega^\dagger
\eea
Here $s_V=\sin\theta_V$, and $c_V=\cos\theta_V$, 
and the numerical factors are normalizatons of mesons expressed 
in terms quark fields. The combinations of constants 
$k_{\bar{q}_a q_b}^{P,V}$ are contained in expressions 
for all quark currents 
(from the scalar-quark current to the tensor-quark current) 
and we introduce them to abbreviate the expressions for box form 
factors.\\
The photon-penguin and the Z-boson--penguin amplitudes 
contain combinations of constants 
$k_{\bar{q}_a q_b}^{P,V}$ given in Table 3. 
For instance, $k_\gamma^{\rho^0} = Q_u k^{\rho_0}_{\bar{u}u} +
Q_d k^{\rho_0}_{\bar{d}d} + Q_s k^{\rho_0}_{\bar{s}s}$.\\
\begin{center}
\begin{minipage}[h]{15cm}
\begin{center}
\begin{tabular}{llllll}
\hline\hline
$k$ & $\rho^0$ & $\phi$ & $\omega$ \\
\hline
$k^{V}_{\gamma}$ &
$\frac{1}{\sqrt{2}}$ &
$\frac{1}{\sqrt{6}}c_V$ &
$\frac{1}{\sqrt{6}}s_V$ \\
$k^{V}_Z$ &
$\frac{1}{\sqrt{2}}c_{2W}$ &
$\frac{c_V}{\sqrt{6}}c_{2W}+\frac{s_V}{2 \sqrt{3}}$ &
$\frac{s_V}{\sqrt{6}}c_{2W}-\frac{c_V}{2 \sqrt{3}}$ \\
\hline\hline
$k$ & $\pi^0$ & $\eta$ & $\eta'$ \\
\hline
$k^{P}_Z$ &
$\frac{1}{\sqrt{2}}$ &
$\frac{c_P}{\sqrt{6}}+\frac{s_P}{2\sqrt{3}}$ &
$\frac{s_P}{\sqrt{6}}-\frac{c_P}{2\sqrt{3}}$ \\
\hline
\end{tabular}\\
\end{center}
{\bf\small Table 3.}\ {\small
Combinations of $k_{\bar{q}_a q_b}^{P,V}$ constants appearing in
photon-penguin and the Z-boson--penguin $\ell \to \ell' P(V)$ amplitudes.} 
\end{minipage}\\[3cm]
\end{center}

%
%

\end{document}